\documentclass[12pt,onecolumn,twoside]{ctt_style/CTT}

\usepackage{graphicx}

\usepackage{multicol}

\usepackage{amsmath}
\usepackage{amsthm}
\usepackage{amsfonts}
\usepackage{amssymb}
\usepackage{mathrsfs}

\usepackage{lettrine}

\usepackage[sort]{cite}

\usepackage{times}

\usepackage{booktabs}

\usepackage{wasysym}

\usepackage{fancyhdr}

\usepackage{ulem}

\usepackage{picins}

\def\ubar#1{\underset{\raise0.3em\hbox{$\smash{\scriptscriptstyle-}$}}{#1}}

\allowdisplaybreaks

\newtheoremstyle{mystyle}{1pt}{1pt}{\normalfont}{\parindent}{\bfseries}{}{1em}{}
\theoremstyle{mystyle}

\renewcommand{\footnotesep}{1pt}
\renewcommand{\footnotesize}{\scriptsize}


\usepackage[utf8]{inputenc}
\usepackage{array}
\usepackage{subcaption}
\usepackage{multirow}
\usepackage{graphics,graphicx} 
\usepackage{mathptmx} 
\usepackage{times} 
\usepackage{psfrag}
\usepackage[dvipsnames]{xcolor}
\usepackage{capt-of}

\begin{document}                          
\makeatletter
\def\@autr{{M. Focchi} et al.}             
\makeatother
\begin{frontmatter}                       

\title{Robot Impedance Control and Passivity Analysis with Inner Torque and Velocity Feedback Loops}
\footnotetext
{\scriptsize $^{\dag}$Corresponding author.
\\
\hspace*{1em}E-mail: michele.focchi@iit.it Tel.: +39 010 71781974; fax: +39 010 71781232.
\\
\hspace*{1em}This work was supported by the Istituto Italiano di Tecnologia.
\\[1mm] %
\hspace*{1em}$\copyright$ 2015 Istituto Italiano di Tecnologia, Advanced Robotics Department, Italy.}

\author[1]{Michele Focchi}$^{\dag}$,{ }
\author[1]{Gustavo A. Medrano-Cerda},{ }
\author[2]{Thiago Boaventura},{ }
\author[1]{Marco Frigerio},{ }
\author[1]{Claudio Semini},{ }
\author[2]{Jonas Buchli},{ }
\author[1]{Darwin G. Caldwell},{ }

\address[1]{. Dept. of Advanced Robotics, Istituto Italiano di Tecnologia (IIT), via Morego, 30, 16163 Genova, Italy;}
\address[2]{. Agile \& Dexterous Robotics Lab,  ETH Z\"urich,  Tannenstr. 3, 8092 Z\"urich, Switzerland;}

\cttdate{Received 21st January  2015; revised 19th October 2015; accepted  22nd  October 2015}

\begin{abstract}

Impedance control is a well-established technique to control interaction forces in robotics.
However, real implementations of impedance control with an inner loop may
suffer from several limitations.  
In particular, the viable range of stable stiffness and damping values
can be strongly affected by the bandwidth of the inner control loops (e.g. a torque loop) 
as well as by the filtering and sampling frequency. 
This paper provides an extensive analysis on how these aspects influence the 
stability region of impedance parameters as well as the passivity of the system. 
This will be supported by both simulations and experimental data. 
Moreover, a methodology for designing joint impedance 
controllers based on an inner torque loop and a positive velocity feedback 
loop will be presented. The goal of the velocity feedback 
is to increase (given the constraints to preserve stability) the  bandwidth of the torque loop 
without the need of a complex controller. 
\end{abstract}


\begin{DOI}
10.1007/s11768-014-****-*
\end{DOI}

\end{frontmatter}

\makeatletter
\renewcommand{\headrulewidth}{0pt}
\fancypagestyle{plain}{\fancyhf{} \fancyhead[LE,LO]{}
\fancyhead[CE,CO]{Control Theory and Technology} \fancyhead[RE,RO]{} \lfoot{} \cfoot{}
\rfoot{}} \thispagestyle{plain}
\pagestyle{fancy}\fancyhf{} \fancyhead[RE]{} \fancyhead[CE]{\small
\hfil{\it\@autr{ / }}\@jou@vol@pag\hfil} 
\fancyhead[CO]{\small \hfil{\it\@autr{ / }}\@jou@vol@pag\hfil}
\fancyhead[LO]{}
\fancyfoot[LE,RO]{}\fancyfoot[RE,LO]{}
\fancyfoot[CE,CO]{} \makeatother

\begin{multicols}{2}                



\section{Introduction}
\label{sec:intro}

Until recently, the majority of legged robots employed high-gain (stiff) position 
feedback control \cite{pratt02}. However, this approach is unsuitable when a robot is in contact with unstructured real-world environment,
as the controller would try to satisfy the position goal at all costs \cite{buchli2009}.
Instead, for such scenarios, a force/torque control in joint or end-effector space is desirable.

For a legged robot, force control can be useful in both the swing
and stance phase. During stance, it allows to control the 
ground impact forces, with the purpose to improve balance capabilities.
During the swing phase, it plays a crucial role in providing to the robot's leg  the compliance necessary to negotiate unperceived obstacles,
 while still ensuring a good position tracking by using  
rigid body inverse dynamics.
%
Interaction forces can be regulated in two ways: passively and actively. Passive methods are those in which physical compliant elements are included between the robot and the environment 
%
%
%
to limit the interaction forces 
%
%
%
(e.g. a passive spring in series elastic actuators  \cite{morfey2013spring}, \cite{Paine2014}). On the other hand \textit{active} compliance is 
achieved through the active control of joints (position or torque) using feedback measurements  
of joint torques \cite{albu-schaffer07}. This can \textit{emulate} a \textit{virtual} compliance both at the joint as well as at the end-effector/foot level.

A major benefit of active compliance is its ability to change
the dynamic characteristics (e.g. stiffness and damping)
\textit{in real-time}. Hence, legged robots can take advantage of active compliance to adapt 
the leg stiffness to swing and stance phases, or to the surface properties
\cite{ferris98LegStiffnessDiffSurfaces}.  
%
Many methods to actively control compliance
at the end-effector have been developed, such as impedance control \cite{hogan85b}, operational
space control \cite{khatib87}, hybrid force-control \cite{raibert81}, and virtual model control 
\cite{Pratt2001}.
Impedance control, in particular, allows the dynamic characteristics at the robot interaction port (e.g. the end-effector) to be specified by regulating the dynamic relationship between forces and positions (mechanical 
impedance). 
Despite impedance is of primary importance to achieve dynamically stable
robot locomotion, only recently  an exhaustive research  has 
been carried out, on the  MIT Cheetah robot, to find 
which impedance parameters are suitable for locomotion 
\cite{bosworth2014effect}.
However, an analysis that investigates if these parameters are 
\textit{realizable} is still missing. 

In the past,  impedance control algorithms were  limited by the controller bandwidth, which 
was set by the computational power and actuator dynamics. That was one of the reasons
for the introduction of passive elements in series with the actuator \cite{Williamson1995}, which have intrinsically unlimited \textit{bandwidth}. However, recent advances in both computer and actuator performance, made active compliance feasible for highly-dynamic applications 
\cite{Boaventura2012a, Hutter2012}.
Nevertheless, many aspects, still create stability issues on impedance control.
For instance, the range of stable stiffness and dampings that can be virtually created (\textit{Z-width} \cite{Mehling2005}, where $Z$ stands for impedance 
\cite{colgate94journal}) can be limited by filtering, sampling frequency, and also
by the bandwidth of inner control loops (e.g. a torque loop).

A common practice in designing nested loop control systems is to maximize the bandwidth of the innermost 
loop \cite{ellis2000}. However, maximizing the inner loop controller bandwidth is not always the 
best strategy. When the outer impedance loop is closed, designing 
the inner loop to have the highest possible bandwidth reduces the 
range of impedance parameters for which the whole system is stable, 
as demonstrated later in this work. Therefore, a trade-off must be found 
between: having a high bandwidth to ensure good 
torque and impedance tracking, and keeping the bandwidth low to increase 
the range of stable impedance values. 
%
Other aspects that directly influence the stability region are the sampling frequency and filtering
\cite{sharifi2000}. Their effect is to introduce delays into the control loop, 
and their influence will 
also be investigated in this work.
%
%
To ensure closed-loop stability during interactions with the 
environment or other systems, the controller  
must be designed to ensure the system behaves 
\textit{passively} at the interaction port \cite{Albu-Schaffer2007},\cite{Schaffer2004}. 
From the passivity property, asymptotic stability can always be ensured: both in free motion as 
well as when the robot is in contact with any type of environment (which is usually
passive). Physical compliant elements and rigid bodies are passive by nature. 
However, when the compliant behavior is emulated by an actuator, 
the passivity is a function of the controller gains.
In this work it will be shown that passivity can also be a  restrictive condition to  
select impedance parameters.

\textit{Related works}. The published literature about active compliance is vast. 
A brief review on the  issues that 
affect the performance of force controlled 
robots can be found in \cite{hogan1989stability}.
Stability analysis and performance specifications for compliance control was first introduced by 
Kazerooni \textit{et al.} \cite{Kazerooni1986} for a manipulator whose model had bounded uncertainty.
Lawrence in \cite{Lawrence1988} considers the non-ideal, 
practical effects of computation and  communication delays 
on impedance control and finds some stability boundaries.
However, his analysis was in continuous time and it 
is not necessarily valid for discrete time systems. Indeed sampling 
is not completely equivalent to time delays because when sampling 
there are additional zeros that do not appear in continuous time.

Regarding controllers based on passivity, 
Albu-Schaffer \textit{et al.} in \cite{Schaffer2004} 
implemented a full state controller for joint or Cartesian impedance with passive capabilities. 
The controller is not passive itself but it is together with the motor dynamics. 
The torque feedback shapes the rotor inertia of the motors to a desired value.  
More recently Buerger and Hogan \cite{Buerger2007} have revisited the problem of designing controllers 
for physically interactive robots. 
For  a 1 DoF system, they reformulated the problem as a robust stability problem 
based on $\mu$-synthesis (structured singular values) and loop shaping methods. The approach provides 
improvements in robot performance compared to traditional passive controllers.
In \cite{Mehdi2011} stiffness and impedance control
concepts were used for robot-aided rehabilitation. 
New stability conditions were proposed using Lyapunov approach and
based on the relationship between the dynamics of the robot
and its energy. 
In \cite{yasrebi2008extending} Yasrebi et al. carried out a time-domain passivity analysis of the Z-width
diagram. This led to the design of a new haptic controller which extended 
the range of stable impedance parameters (Z-
width) by means of an acceleration feedback. 
The analysis was carried out for one joint 
using passivity theory in the frequency domain. 

The main contribution of this work, 
is a methodology to analyze (based on an
accurate model)  stability and passivity  
of a gearbox driven actuator (plus load) system.
The analysis takes into account all the non-idealities 
present in \textit{real} implementation of an \textit{impedance} 
controller, namely: actuator dynamics, discrete implementation, filtering, nested loops. 
This allows to find the impedance \textit{"stability regions"} 
which represent the impedance parameters that 
can be \textit{rendered} in a \textit{stable} way.
Simulations and experimental data show how the above-mentioned 
non-idealities influence the \textit{stability regions} as well 
as the passivity of the system.
The study is carried out for the adduction/adduction (\textit{HAA}) 
joint of the HyQ \cite{semini01092011} robot (see Fig. \ref{fig:hyq}), 
where impedance control was implemented with an inner torque loop \cite{amc12focchi}.
However, the underlying ideas are \textit{valid} for any electric actuator moving a load with a gearbox reduction. 
In the bigger picture, the \textit{stability regions} 
are the basis to develop a gain scheduler (in the low-level control layer) 
which is able to adapt the bandwidth of the inner torque loop 
according to the impedance parameters set by the user.
\begin{center}
\centering
  \includegraphics[width=1.0\columnwidth]{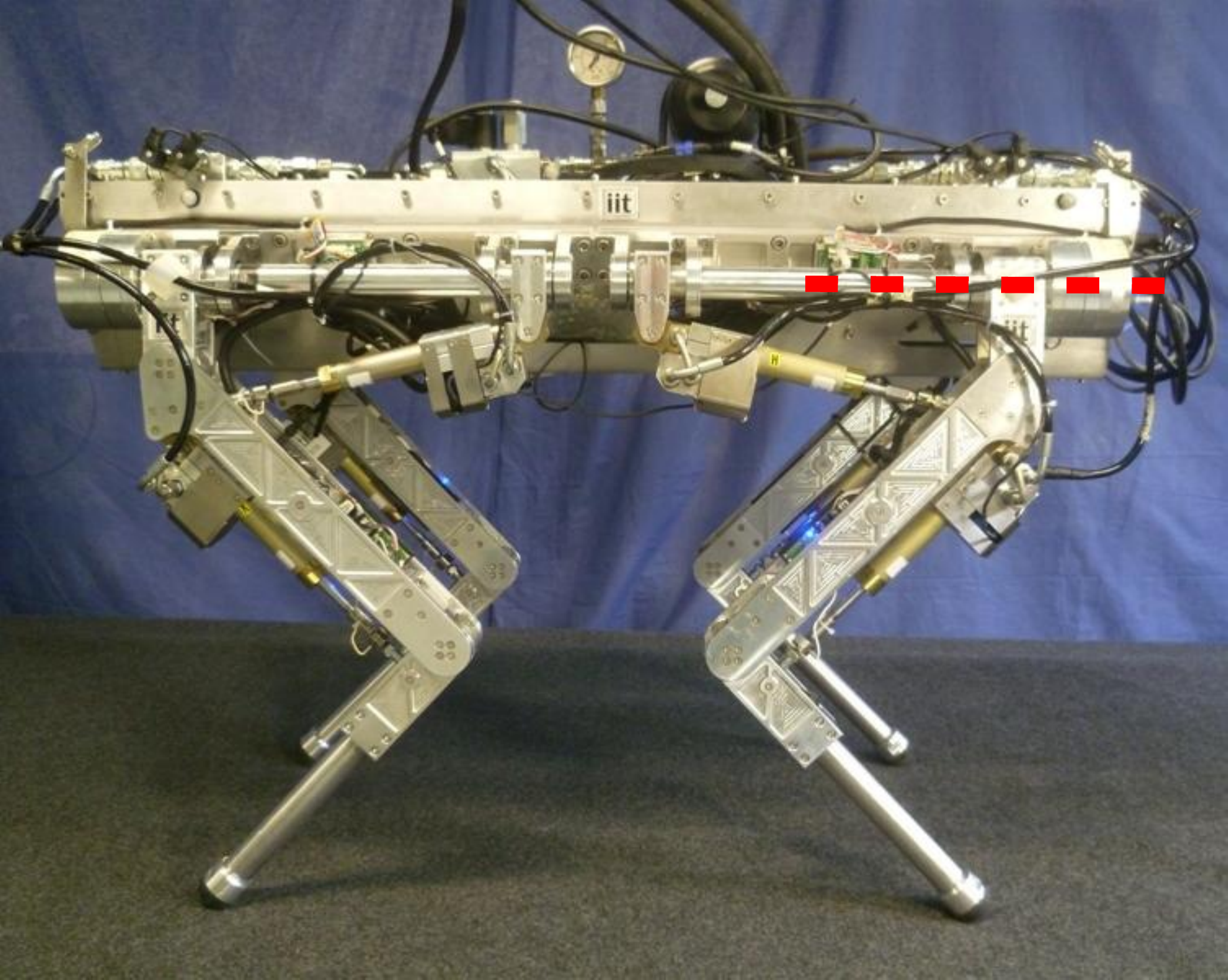}                
  \figurecaption{Picture of a lateral view of the HyQ robot, the HAA joint axis is marked in red.}
  \label{fig:hyq}
\end{center}
This paper is structured as follows: the mathematical model of the system is introduced in Section \ref{sec:model} 
followed by a description of the control system implementation in Section \ref{sec:controller}. 
The stability issues associated with real implementation 
of an impedance controller are 
analysed, both in simulations and experimentally, in Section \ref{sec:simExp}. 
A brief assessment about passivity for 
the system is then given in 
Section  \ref{sec:passivity}.
Finally, Section \ref{sec:conclusions} discusses the results and future works.


\section{System description and mathematical model}
\label{sec:model}
The studies and experiments presented in this work are all conducted on
HyQ \cite{semini01092011}.
HyQ is a fully torque-controlled quadruped robot with a mix of hydraulic and electric actuation 
for each leg: two hydraulic joints on the 
sagittal plane (hip HFE and knee KFE flexion-extension) and one electric joint moving 
in the traversal plane (hip adduction-abduction HAA,  Fig.\ref{fig:leg_sketch} on the  left). 
This paper focuses on modeling and control of the electric joint, which 
consists of a DC 
brush-less motor (Emoteq HT2301) and a harmonic drive gearbox (CSD-25-100). 
The leg is attached to the gearbox output via an interface consisting of 6 parallel 
pins (evenly distributed on a circle around the axis of rotation) that
enable easy dismounting (see Fig. \ref{fig:motor_assembly}). This interface represents a small intermediate rotational inertia ($J_{L1}$) placed  before the inertia represented by the leg ($J_{L2}$) in the transmission train.

\begin{center}
\centering  
\includegraphics[width=0.47\columnwidth]{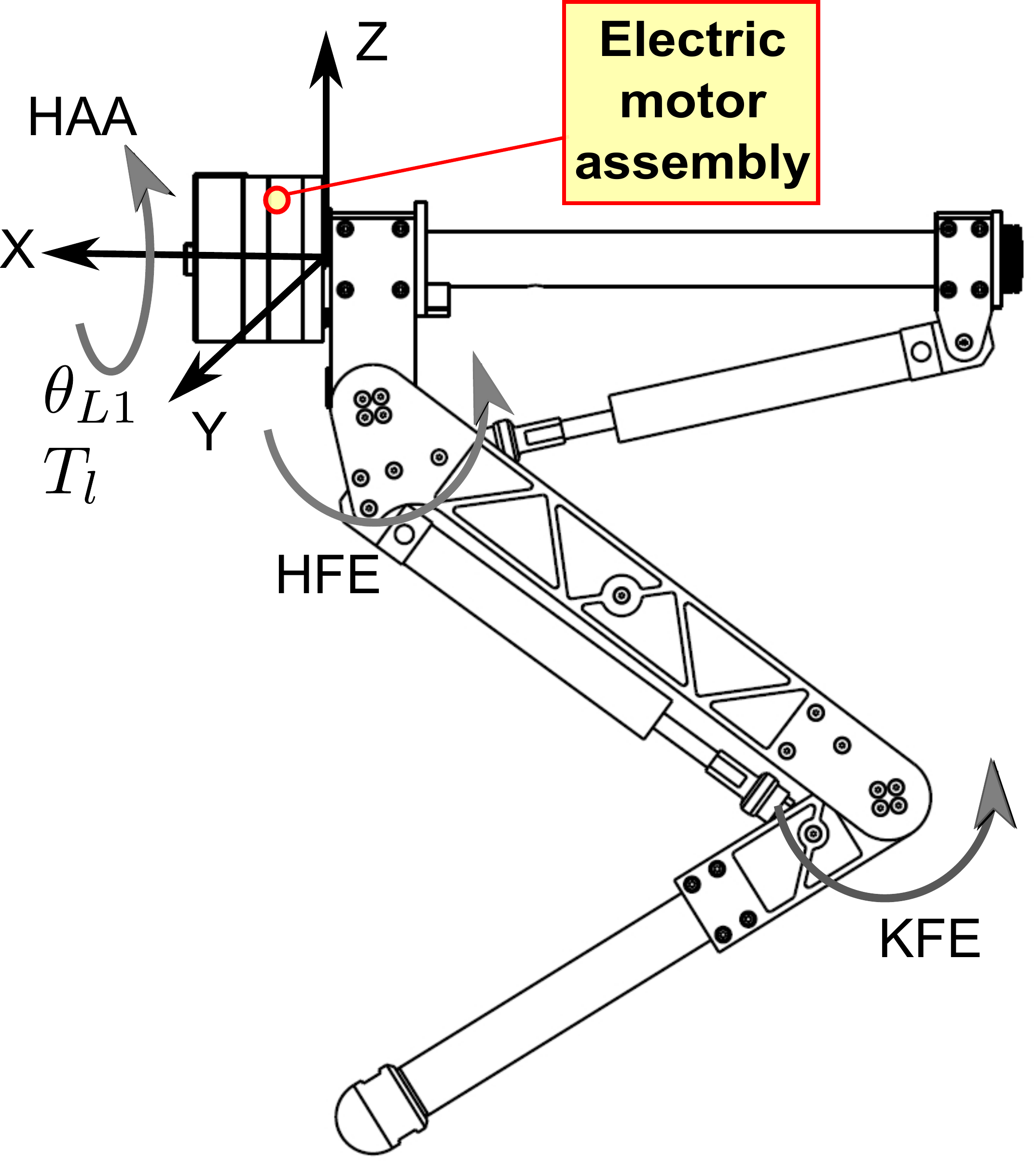}            
\includegraphics[width=0.47\columnwidth]{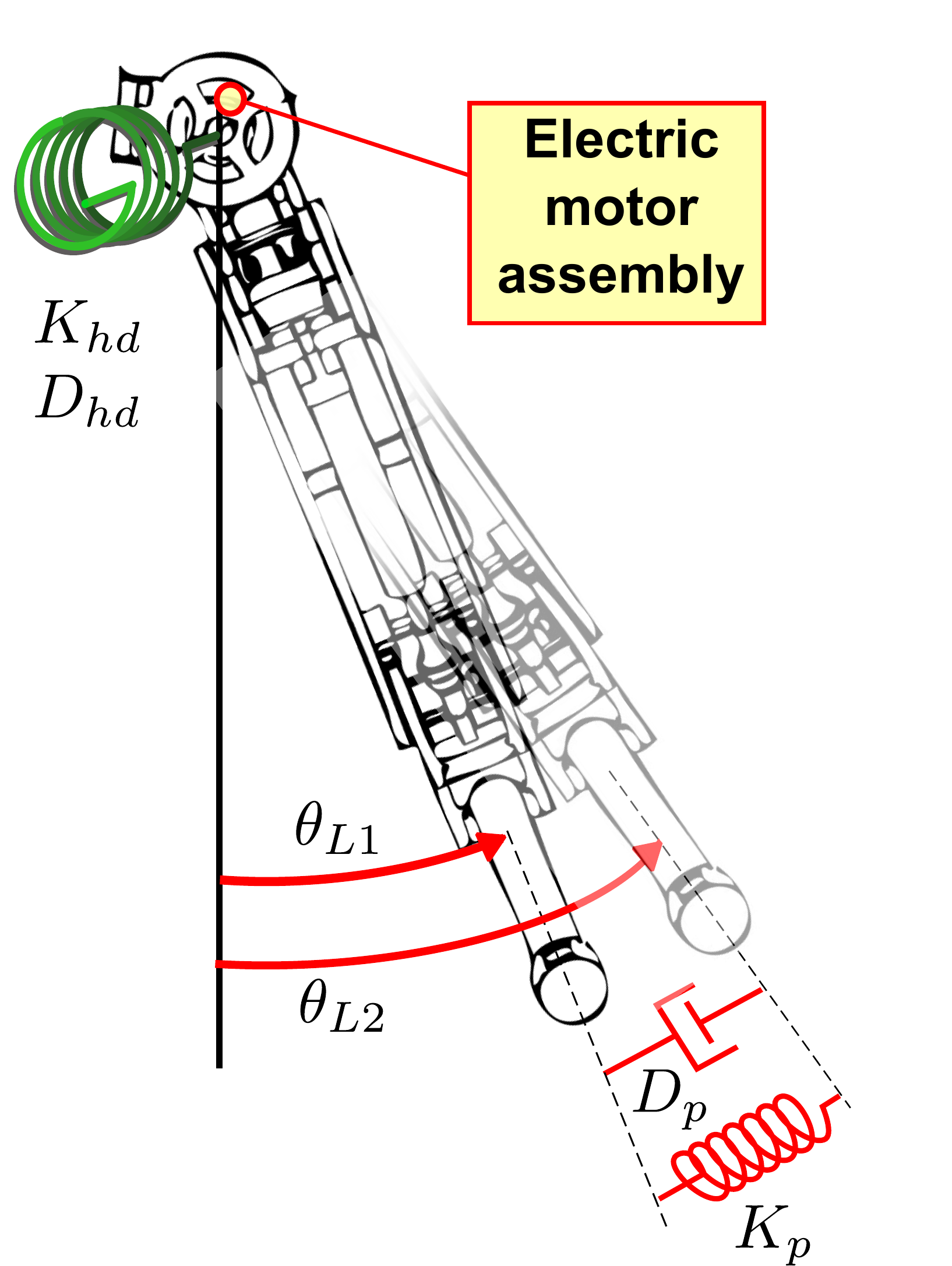}
\captionof{figure}{HyQ Leg. Lateral (\textit{left}) and frontal (\textit{right}) view. The figures show the definition  of the joints and their angles, as well as the coordinate frame.}
\label{fig:leg_sketch}
\end{center}

Normally, this type of assembly is modelled by two second order 
differential equations coupled via the gearbox transmission flexibility \cite{hori94}. 
However, after performing several open loop tests using chirp signals, 
an anti-resonance was detected for the link velocity (see the frequency response of the link velocity to a chirp input voltage in Fig. \ref{fig:TFlinkvel}).

\begin{center}
\centering
	\includegraphics[width=0.8\columnwidth]{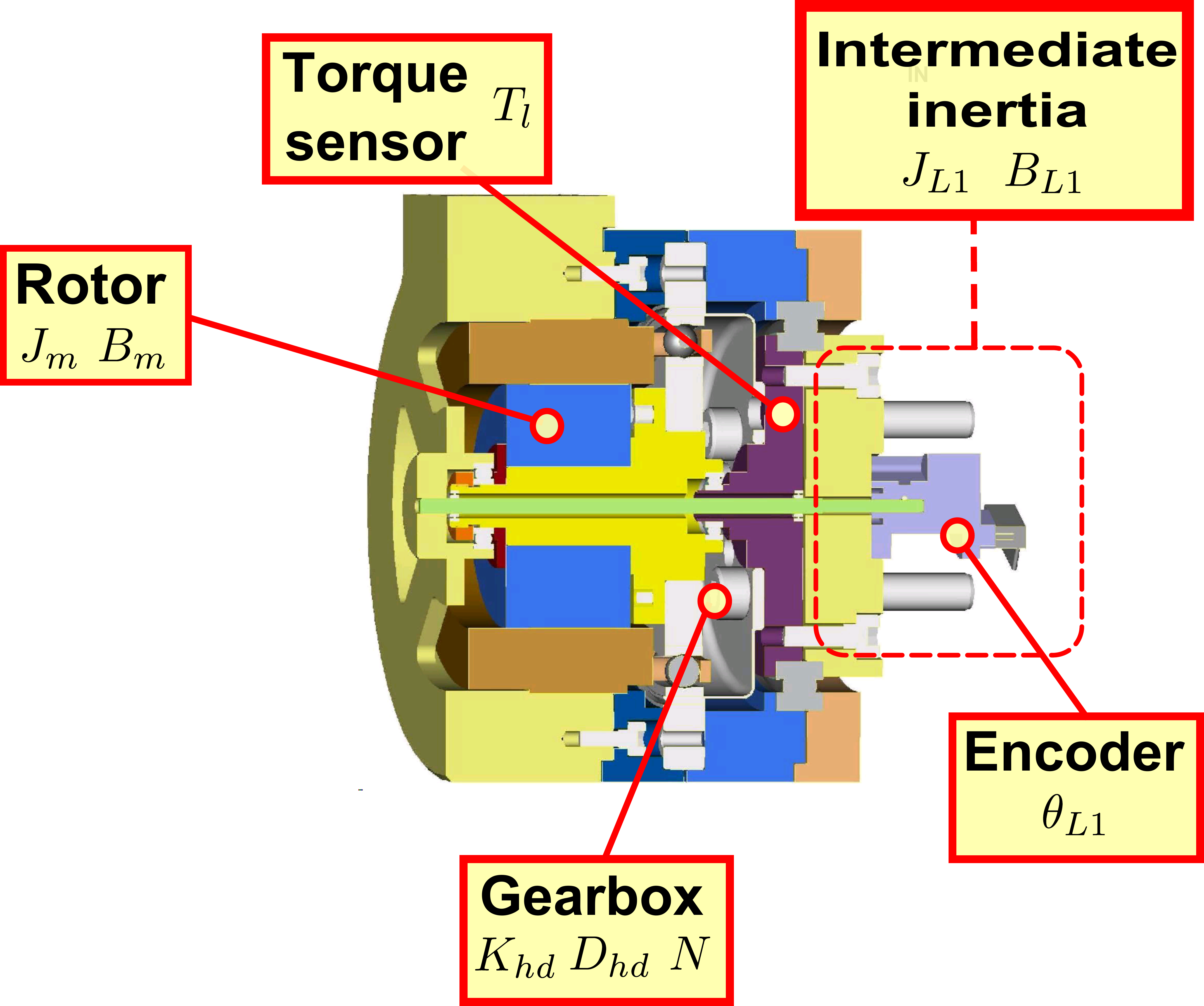}
	\figurecaption{Cross-section of the mechanical assembly of the electric joint. 
                        The intermediate inertia $J_{L1}$ 
		represents the part that interfaces the leg with the gearbox output.}
		\label{fig:motor_assembly}
\end{center}

Since a model with two inertia and one spring cannot capture this behavior, a more 
complex model with three inertia coupled by springs was used, as shown 
in the schematic in Fig. \ref{fig:motorscheme} where $K_{hd}$ and $K_p$ are 
the stiffness related to the gearbox and the leg flexibility, respectively.

\begin{center}
  \centering
	\includegraphics[width=0.9\columnwidth]{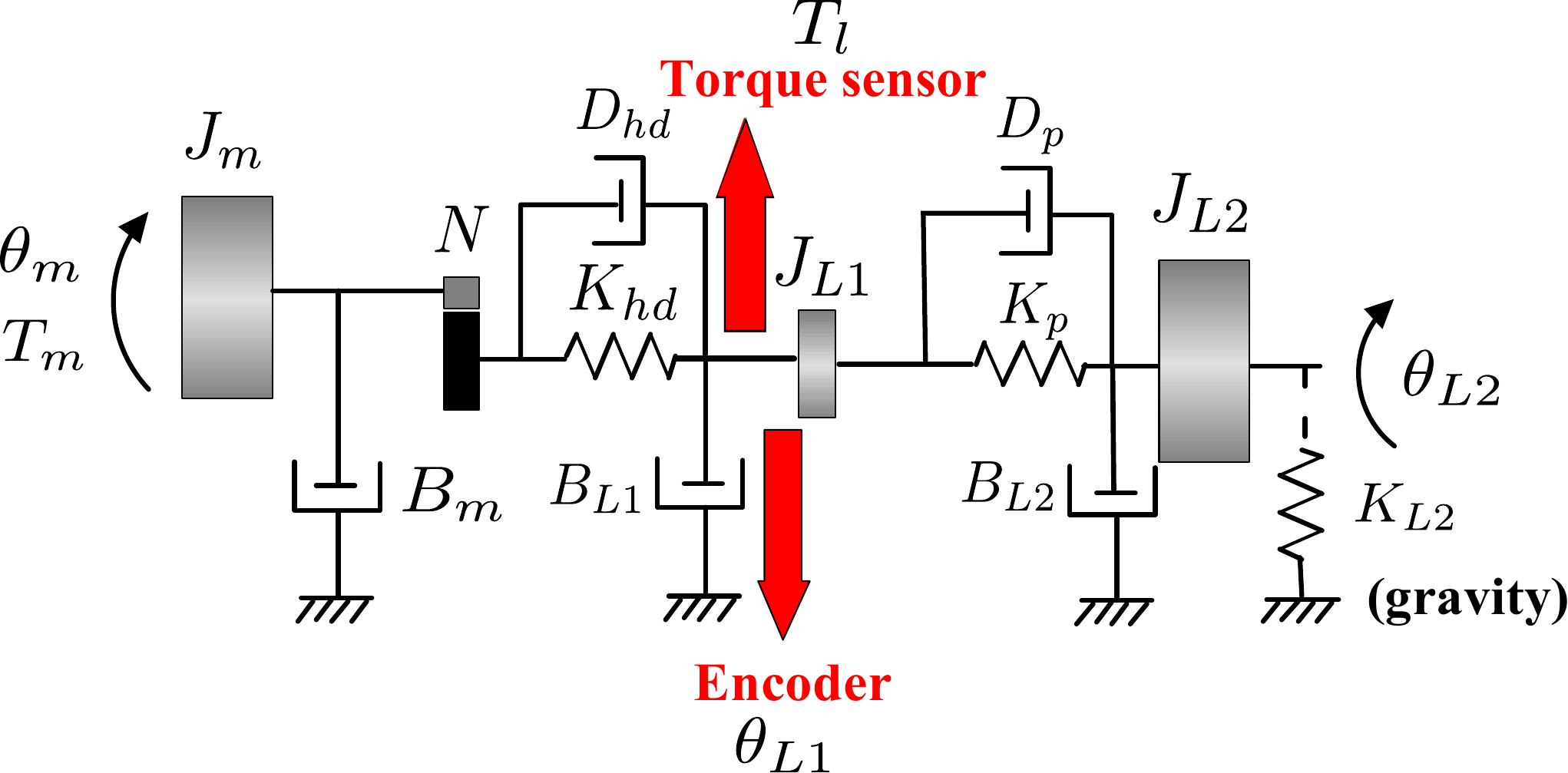}
	\figurecaption{Motor drive system with torsional load, schematic for the 3 mass-2 spring model.}
	\label{fig:motorscheme}
\end{center}
Joint position and torque are measured by an encoder and a torque sensor.
Due to the topology depicted in Fig. \ref{fig:motor_assembly} the position encoder measures the 
angle of the intermediate inertia $J_{L1}$ while the link velocity is measured by averaging 
first order differences (4 samples) of the position encoder (averaging
filter). In addition a strain gage based torque sensor is mounted at the output of 
the harmonic drive. No filter is implemented on the torque signal because it would introduce delays in the control action. 
According to this model, the Laplace transforms of the differential equations 
that describe the linearized dynamics of the load and of the electric motor are:

\small	
\begin{equation}
\renewcommand{\arraystretch}{2.5}
\begin{array} {l}
I_m = \dfrac{V_m}{Ls+R}-\dfrac{k_ws \theta_m}{Ls+R} \\                                                                                               
\theta_m = 	\dfrac{k_t I_m}{(J_m s+B_m)s} -\dfrac{K_{hd}+sD_{hd}}{N(J_m s+B_m)s}\left(\dfrac{\theta_m}{N}-\theta_{L1}\right) + T_{fr}\\                           
\theta_{L1} = \dfrac{-(K_p+sD_p)}{(J_{L1} s+B_{L1}) s}(\theta_{L1}-\theta_{L2})  +\dfrac{K_{hd}+sD_{hd}}{(J_{L1} s+B_{L1}) s} \left(\dfrac{\theta_m}{N}-\theta_{L1}\right)   \\
\theta_{L2} = \dfrac{K_p+sD_p}{(J_{L2} s+B_{L2}) s}\left(\theta_{L1}-\theta_{L2}\right)- \dfrac{K_{L2}\theta_{L2}}{(J_{L2} s+B_{L2}) s} +T_{dist}\\                                    
\end{array}
\label{eq:motorloadDyn} 
\end{equation}
\normalsize
\begin{center}
\centering
	\includegraphics[width=0.9\columnwidth]{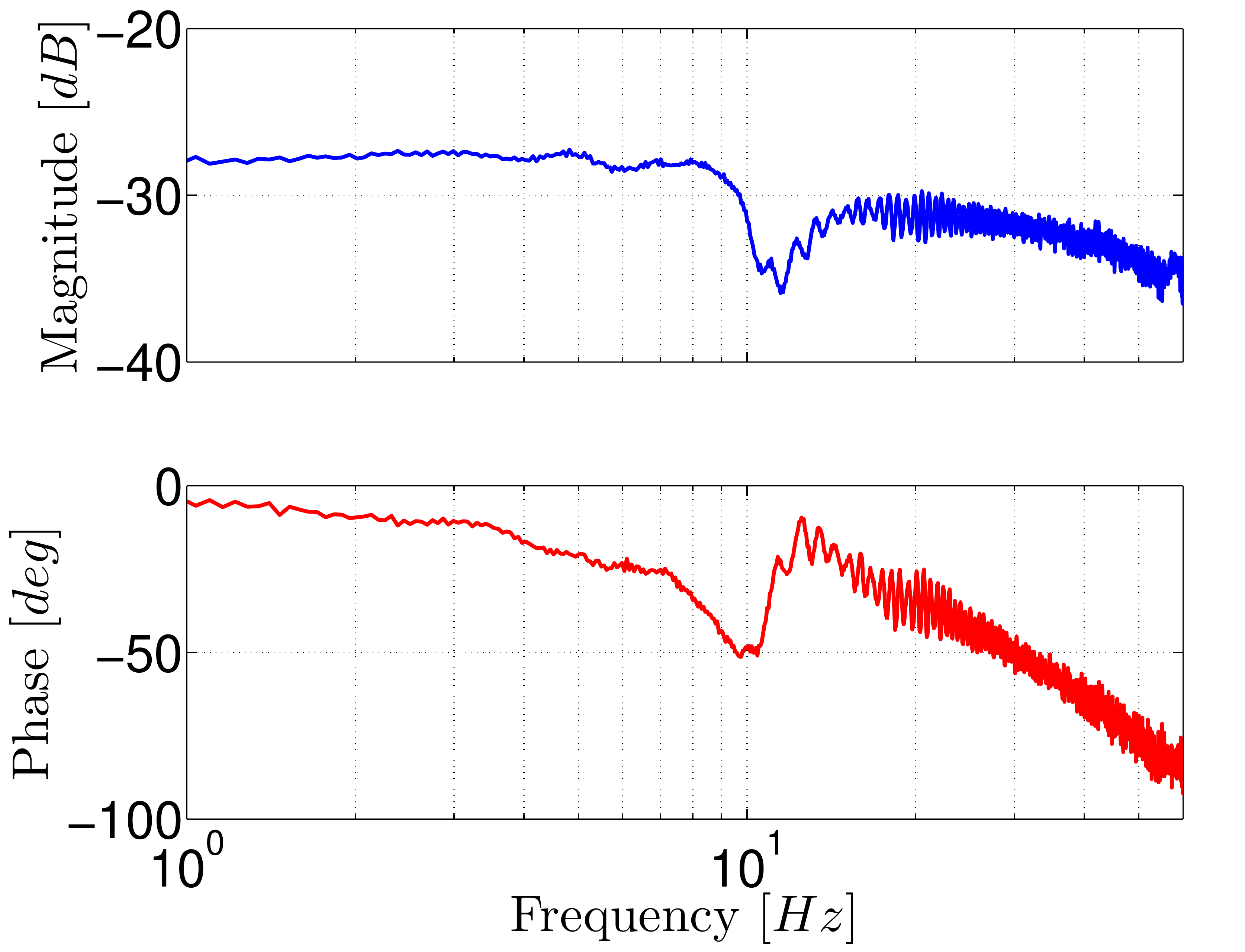}
	\figurecaption{Frequency response of the link velocity $\dot{\theta}_{L1}$ to a chirp input voltage at the HAA motor, experimentally
	obtained with an unconstrained HyQ leg.}
	\label{fig:TFlinkvel}
\end{center}

\begin{figure*}
  \centering
	\includegraphics[width=1\textwidth]{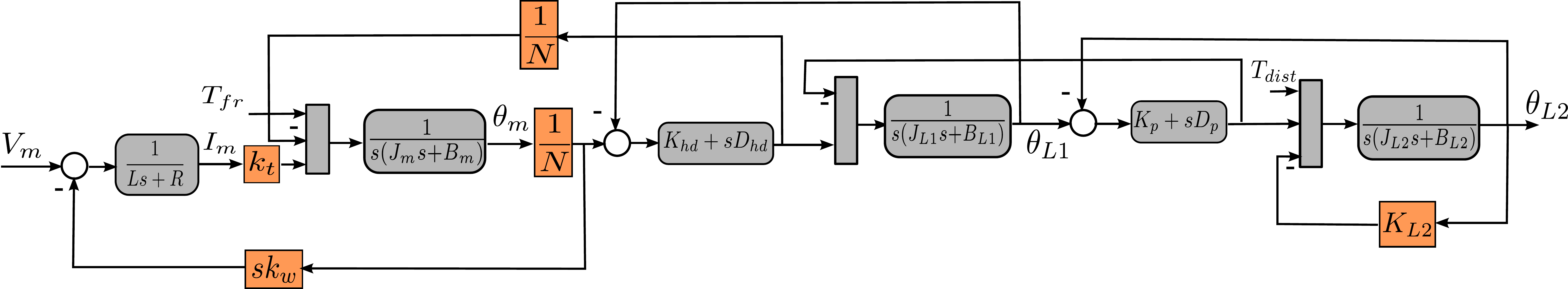}
	\figurecaption{Block diagram representing Eqn. (\ref{eq:motorloadDyn}) that describes 
	the linearized dynamics of the electric motor and of the load. $V_m$ is the voltage input, $T_{fr}$ the harmonic-drive disturbance torque, $T_{dist}$ is an external disturbance torque coming from the load side, $\theta_m$ is the motor position, $\theta_{L1}$ and $\theta_{L2}$ the positions of the intermediate and leg inertia, respectively.}
	\label{fig:block_diagram_equation}
\end{figure*}

where $I_m$, $\theta_m$   denote the motor current and motor position; $\theta_{L1}$  and 
$\theta_{L2}$   are the intermediate inertia and leg positions;  $V_m$ is the motor 
voltage; $T_{fr}$  is the friction torque in the harmonic drive and  $T_{dist}$ 
is an external disturbance torque applied to the leg. All other symbols and parameter 
values are given in Table  \ref{tab:params}. 
A block diagram that illustrates the relationships between the state variables is depicted in 
Fig. \ref{fig:block_diagram_equation}.
Since the rotational inertia of the leg $J_{L2}$
varies with the configuration of the joints HFE and KFE, the table includes also 
upper and lower bounds for the leg inertia. In particular the fact that the leg can 
retract or extend results in different mass distributions around the hip abduction-adduction axis.
Namely, the inertia is higher when the leg is extended and lower when it is retracted.
Gravity has also an effect on the load dynamics, that, when linearized, behaves
as a \textit{rotational} spring $K_{L2}$. The stiffness of this ''gravity spring'' is also dependent on the leg configuration
and is reported in Table \ref{tab:params} for an extended and retracted leg configuration.
%

%
\begin{center}
\centering
\captionof{table}{Model parameters}
\label{tab:params}
\renewcommand{\arraystretch}{1.1}
\begin{tabular}{p{0.7cm} p{4cm} p{2.7cm} }
\hline \hline
\textbf{Name} 			& 		\textbf{Model Parameters}	& 		\textbf{Value} 	     					    \\
\hline
\hline								
$J_{m}$						&    Rotor + gearbox inertia  (D)		&5.72 $\cdot 10^{-5} kgm^2$							\\
\hline	
$K_{hd}$						&	Gearbox stiffness (I) &8.077 $\cdot 10^3  Nm/rad$  										\\
\hline
$D_{hd}$						&	Gearbox damping   (I) &16.56	 $Nms/rad$												\\
\hline
$B_m$						&	 Visc. frict. rotor (I) & 0.0015 $Nms/rad$												\\
\hline
$J_{L1}$					    &	 Interm. inertia   	(I)	& $1\cdot10^{-4}$ $kgm^2$				 					\\
\hline
$B_{L1}$					    &    Visc. frict. of inertia $J_{L1}$ (I)& 0  $Nms/rad	$									\\
\hline
\multirow{2}{*}{$J_{L2}$}  	&	\multirow{2}{*} {Leg inertia (C)} 	& 0.439 $kgm^2$ (ext.) 						   	\\	
							&	 							& 0.129 $kgm^2$ (ret.)  							    \\	
\hline
$B_{L2}$					    &    Visc. frict. of  inertia   $J_{L2}$ (I)   & 0.756  $Nms/rad$								\\	
\hline
\multirow{2}{*}{$K_{L2}$}	&    Linear stiffness 			&11.2 $Nm/rad$ 		 (ext.)								\\			
				 			&	due to gravity			    &7.17 $Nm/rad$ 		 (ret.) 							\\
\hline				 			
$K_{p}$						&    Leg stiffness  (I)            & 1.923$\cdot 10^3 Nm/rad$								\\	
\hline
$D_p$						&    Leg damping (I) 				& 7.56 $Nms/rad	$										\\
\hline
$L$							& 	Coil inductance  (D)&  2.02$\cdot 10^{-3} H$											\\
\hline
$R$							&  	Coil resistance (D) & 3.32 $\Omega $														\\
\hline
$k_t$						&   Motor torque constant (D)&0.19 	$Nm/A$													\\
\hline
$k_w$						&   Motor speed constant (D) &0.19 $Nms/rad$						    						\\
\hline
$N$ 						&   Gear ratio (D) & 100 											   						\\
\hline\hline
\end{tabular}
\begin{tabular}{p{0.7cm} p{5.5cm} p{1.2cm} }
\textbf{Name} 	& 		\textbf{State variables}	& 		      \textbf{Unit}         	  					\\
\hline\hline
$\theta_m$					&	Ang. pos. of the rotor & $rad$	 		\\
\hline
$\theta_{L1}$ 				& 	Ang. pos. of the intermediate inertia & $rad$  				\\
\hline
$\theta_{L2}$			& Ang. pos. of the leg & 						$rad$					\\
\hline
$I_m$					& Motor current & $A$	 													\\	
\hline \hline
\textbf{Name} & 		\textbf{Inputs}	& 		\textbf{Unit} 	         		   	    \\
\hline \hline
$V_m$					& Motor voltage & 							$V$	 					  	\\
\hline
$V_{m_{VC}}$  			& Vel. comp. voltage 				&	$V$							\\
\hline
$T_{dist}$ 				& Ext. dist. torque (load side) 	 & 	$Nm$								\\
\hline
$T_{fr}$ 				& Frict. dist. torque (motor side) &		 $Nm$								 \\
\hline\hline
\textbf{Name} 	& 		\textbf{Outputs}	& 		\textbf{Unit} 	            			  \\
\hline\hline
$T_l$   					& Load torque 	 				& 	$Nm$										\\
\hline\hline
\textbf{Name} 	& 		\textbf{Controller Gains}	 	&	 	            	  					\\
\hline\hline
$P_t$					& Torque controller prop. gain & 										\\
\hline
$I_t$				& Torque controller integral gain 												& \\
\hline
$\beta$ & Gain of the PI torque controller 															& \\
\hline
$\alpha$  & Velocity compensation (scalar) gain 														& \\
\hline\hline
\textbf{Name} 	& 		\textbf{Transfer functions}	& 		 	            	  \\
\hline\hline
$PI_t(z)$			& PI torque controller & 			\\
\hline
$G_t(s)$ 		& TF between  $V_m$ and $T_l$ & \\
\hline
$VC_{gain}$ 		& TF of velocity compensation  & 		\\
\hline
$G_{t_{VC}}(s)$ & 	TF between $V_m$ and $T_l$  after vel. comp. & \\
\hline\hline
\end{tabular}
\end{center}

\section{Controller Design}
\label{sec:controller}

This section explains the design of the control system. The controller architecture is shown in Fig.
\ref{fig:blockDiagram} where an inner positive velocity 
feedback  loop is followed by a torque loop controller 
and finally an outer impedance (position) loop. Specifications for the impedance loop 
vary depending on the gait, for example a trotting gait frequency is around 2 
Hz for HyQ that has a mass of 75 $kg$. The specifications for the performance will depend on 
the type of locomotion gait and the gains of the impedance loop will vary in a 
specified range. The inner torque loop and the velocity feedback loop must be designed to 
be consistent with these requirements.

\subsection{Positive feedback velocity compensation}
\label{subsec:VC}

One difficulty in the design of the torque loop controller is that the load dynamics may introduce severe limitations in the closed loop performance of the torque loop. This problem has been largely overlooked since in many cases the load dynamics are ignored in the analysis. In this subsection a positive velocity feedback (\textit{velocity compensation}) is introduced to address these limitations and improve the torque bandwidth. To exhibit the above-mentioned limitations, first of all, the system response has been considered after closing the inner velocity feedback loop. The torque transmitted to the load is 
measured by the torque sensor and can be expressed as:
\begin{equation}
		T_l = (K_{hd}+sD_{hd})\left(\dfrac {\theta_m}{N} - \theta_{L1}\right)
		\label{eq:torque}
\end{equation}

\begin{figure*}
  \centering
	\includegraphics[width=0.8\textwidth]{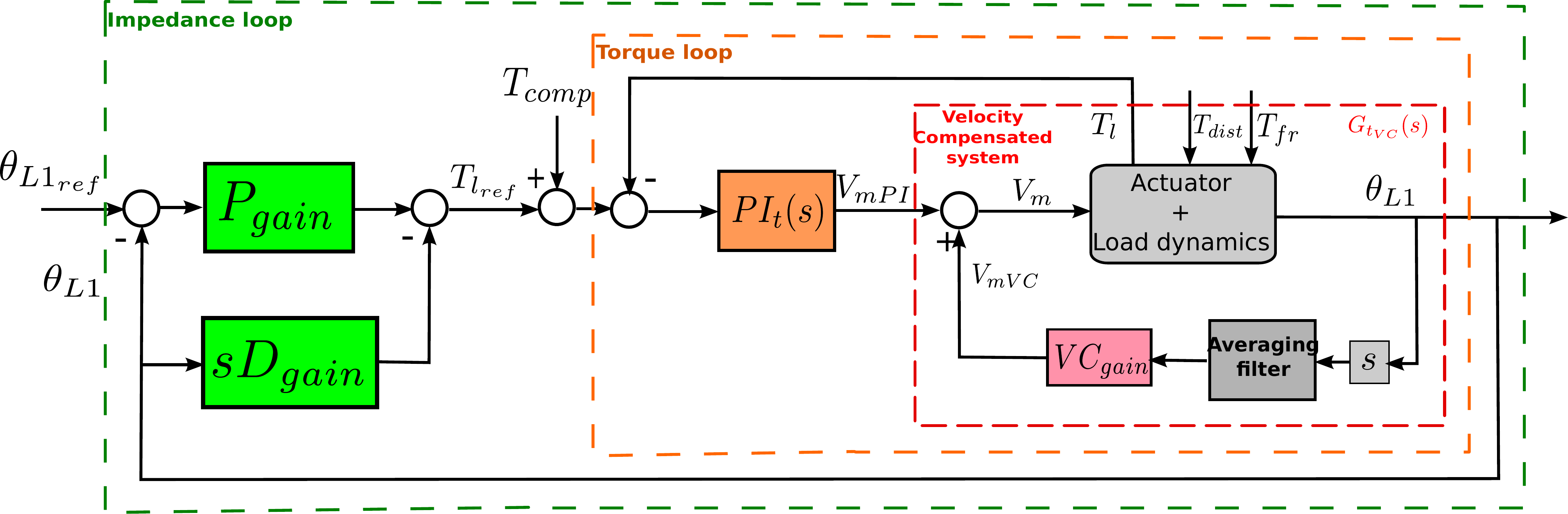}
	\figurecaption{Block diagram of the velocity compensated system with inner torque loop (PI, orange block) outer impedance loop (PD, green blocks). The velocity compensation term ($V_{mVC}$) is added to the output ($V_ {mPI}$) of the torque controller.}
	\label{fig:blockDiagram}
\end{figure*}

From Eqn. (\ref{eq:motorloadDyn}), Eqn. (\ref{eq:torque}) and Fig. 
\ref{fig:blockDiagram}, closing the velocity loop (without considering the 
averaging filter for the sake of simplicity), the transfer function from 
the motor voltage to the torque is given by (cf.  Eq. 3.24, Ch. 4, p. 45 in \cite{focchi13PhDthesis} 
for complete derivation):

\begin{equation}
G_{t_{VC}} = \dfrac{k_t (D_{hd}s + K_{hd})p_1}   {N(p_1 q_1 + (q_2 - VC_{gain}s)q_3)}
\label{eq:torqueTFvc}
\end{equation}
where:
\begin{equation}
\renewcommand{\arraystretch}{1.5}
\begin{array}{l}
{p_1} = ({J_{L2}}{s^2} + {B_{L2}}s + {K_{L2}})({J_{L1}}{s^2} + {B_{L1}}s)\nonumber\\
\qquad + \left[ {({J_{L2}}{s^2} + {B_{L2}}s + {K_{L2}}) + ({J_{L1}}{s^2} + {B_{L1}}s)} \right]({D_p}s + {K_p})
\\
{q_1} = (Ls + R)\left( J_ms^2 + {B_m}s + \dfrac{(D_{hd}s + K_{hd})}{N^2} \right) + {k_t}{k_w}s\\

{q_2} = \dfrac{N}{k_t}(Ls + R)({J_m}{s^2} + {B_m}s) + N{k_w}s\\

{q_3} = \dfrac{k_t}{N}({J_{L2}}{s^2} + {B_{L2}}s + {D_p}s + {K_p} + {K_{L2}})( D_{hd}s + {K_{hd}})\\

\end{array}
\label{eq:torqueTFvc_elements}
\end{equation}


Observe that the transmission zeros in Eqn. (\ref{eq:torqueTFvc}) introduced by the polynomial $p_1$ 
depend entirely on the load dynamics, that is the load connected at the output of the harmonic 
drive. When the damping coefficients $B_{L1}$ and $B_{L2}$  are small, as is usually the case, 
some of these transmission zeros are very close to the stability region boundary. In the case 
of continuous time systems the boundary is the imaginary axis and in the case of discrete time 
systems this is the unit circle. Notice that the zeros may be real or complex depending on 
the value of the stiffness $K_{L2}$. These zeros impose limitations in the achievable closed loop 
bandwidth when using a simple proportional and integral torque controller. This is because the 
controller pole, located at the origin, will be attracted towards the transmission zeros 
becoming the dominant pole of the system, thus limiting the closed loop bandwidth 
of the torque loop unless very high gains are used in the torque controller. 
In most cases the torque loop gain will have a finite gain margin and therefore the controller 
gain cannot be made sufficiently large. This is more pronounced in digital control where the 
gain margin is likely to be much lower than the gain margin for a continuous time system. The 
effect of the velocity compensation is that these unwanted transmission zeros (polynomial $p_1$) 
can be cancelled if the velocity feedback gain is chosen as $VC_{gain}=q_2/s$ so that the term 
$q_2-VC_{gain}s$ in the denominator of Eq. (\ref{eq:torqueTFvc}) is equal to zero. 

\begin{equation}
VC_{gain} = \dfrac{ N}{k_t}(Ls+R)(J_ms + B_m) + Nk_w
\label{eq:vc_gain_full}
\end{equation}

The implementation of the compensator requires 
derivatives of the velocity signal that
is often prone to quantization errors. Since this derivative is likely to be noisy 
the compensator has to be approximated by adding suitable filters which would add delay. 
An alternative solution is to use a simplified velocity compensation 
as presented in  \cite{boaventura12IROS} which is obtained by discarding the 
derivative terms from Eq. (\ref{eq:vc_gain_full}):

\begin{equation}
VC_{gain} = \dfrac{\alpha N}{k_t}(RB_m + k_tk_w) \quad \quad \alpha  > 0
\label{eq:vc_gain}
\end{equation}

This simplified velocity compensation is obtained by setting $L=0$ and $J_m=0$. This means we are
neglecting both the electrical dynamics and the acceleration term, 
which would introduce noise in the system. 
$\alpha$ is introduced as an adjustable parameter. 
Therefore, with Eq. (\ref{eq:vc_gain}), an exact cancellation of the transmission 
zeros  $p_1$ is generally not  possible. Nevertheless, even though an exact cancellation is not 
possible,  an improvement in the closed loop torque bandwidth can be achieved. 
To understand when and how this is 
possible, consider the velocity compensation given by Eq. (\ref{eq:vc_gain}).
For the parameter values given in Table \ref{tab:params} and setting 
$K_{L2}=0$, the polynomial  $p_1$ has four real roots $z_4<z_3<z_2<z_1=0$. 
The transfer function (\ref{eq:torqueTFvc}) also has a pole at zero. 
This is an unobservable pole and therefore it cancels out with the zero 
$z_1$. The second zero $z_2$  is the closest to the imaginary axis and 
limits the torque bandwidth that can be achieved with a PI torque controller. 
Indeed, as  $\alpha$ increases one real pole in Eq. (\ref{eq:torqueTFvc}) moves towards the 
stability boundary along the real axis and at some point it will become identical to $z_2$. 
The value of the gain $\alpha$ for which this happens, is the ideal value 
required for cancelling out unwanted zero $z_2$. In this particular case, $z_2$  
varies as a function of the leg inertia. Hence, it may be difficult to completely 
cancel out this zero for all leg configurations with a fixed value for $\alpha$. However, 
as long as the pole in Eq. (\ref{eq:torqueTFvc})  is placed to the right of the zero $z_2$ 
then the bandwidth limitation introduced by this zero is avoided (cf. Sec. 3.5.3, Ch. 4, p. 46 in \cite{focchi13PhDthesis}). 
When  $K_{L2}\neq0$ then the roots of  $p_1$, which are closest 
to the imaginary axis, are complex ($z_1$ and $z_2$  
are complex conjugate). As $\alpha$  increases two poles in Eq. (\ref{eq:torqueTFvc})  
will move towards the imaginary axis as a complex conjugate pair but there is no value 
of $\alpha$  that will completely cancel out the unwanted zeros $z_1$  and $z_2$. 
In this case the velocity compensation will not be as effective as for the case where  
$K_{L2}=0$. Nevertheless, it still results in an  improvement of the achievable closed 
loop torque bandwidth as shown in Fig. \ref{fig:polesVC_KL2neq0}.  

\begin{center}[th]
  \centering
	\includegraphics[width=0.5\textwidth]{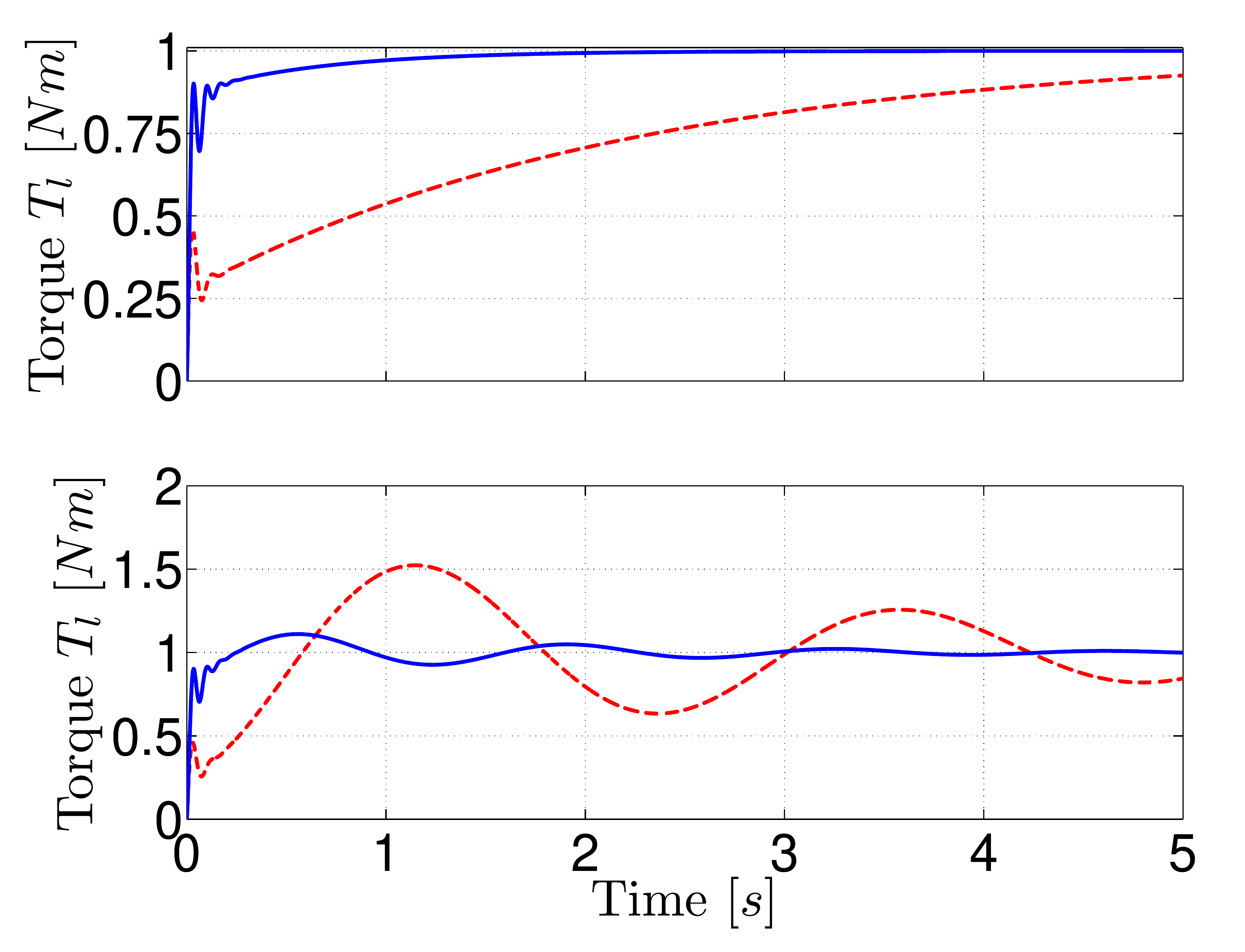}
	\figurecaption{Simulation. Unit torque step responses with $K_{L2}=0$ (\textit{upper plot)} 
	and $K_{L2}=11.2Nm/rad$ (\textit{lower plot)}  for different velocity 
	compensation gains $\alpha=0.94$ (\textit{Solid line}) and $\alpha=0$ (\textit{dashed line}). 
	An extended leg inertia ($J_{L2}=0.439$ $kgm^2$) has been considered. The use of velocity compensation
	significantly reduces the response time.}
	\label{fig:polesVC_KL2neq0}
\end{center}

Similar results can be established after discretization of the 
continuous time system. A final remark is that an alternative to overcome 
the limitations imposed by the unwanted zeros is to consider more complex controllers, 
for example a double integrator PII controller, or designs based on loop shaping 
control which introduce additional poles and zeros. The disadvantage is that tuning 
these complex controllers cannot be done without a good model for the system.

\subsection{Torque controller design}
In this section the design of a PI torque controller is considered. 
An integrator is included in the controller structure to 
remove steady state errors when tracking constant 
torque inputs or when constant disturbances arise. 
The integrator is implemented in discrete time using
the backward euler approximation. This controller is a lag compensator that has one zero and one pole. 
The analysis and design are carried out in discrete time because sampling 
introduces noticeable differences in comparison to the continuous time case. 
For example the system can become non-minimum phase even if the underlying 
continuous time is of minimum phase (non-minimum phase systems are more difficult 
to control). The equation of the PI controller is given by:

\begin{equation}
PI_t(z)= P_t+I_t\dfrac{zT_s}{z-1}=(P_t+I_tT_s)\dfrac{\left(z-\dfrac{P_t}{P_t+I_tT_s}\right)}{z-1}
\label{PItorque}
\end{equation}

where $z$  is the $Z$-transform variable and  $T_s =$ 1 $ms$ is the sampling time interval;  $P_t$ and $I_t$  are the proportional and integral gains to be determined. 
We remark that small sampling time intervals, on one hand, improve the disturbance rejection properties of the closed loop system. However, as the sampling time interval decreases, the effects of quantization noise in the encoders become more prominent, especially when computing velocities from position measurements using simple first order differences. In addition, small sampling time intervals can introduce non-minimum phase behavior in the sampled system which is more difficult to control. The selected sampling time of 1 $ms$ is a trade-off among all these aspects. At the actual encoder resolution (80000 count/rev), the smallest velocity that can be measured with 1 $ms$ sampling time interval, is 0.0125 $rad/s$.

Traditionally, the design of an inner loop controller is carried out with the aim of maximizing the closed loop bandwidth of the inner loop. However, one of the first difficulties is 
how to measure the bandwidth of the torque loop. The closed loop 
torque bandwidth for an unconstrained system,
for example when the leg is moving freely in the air, 
is very different from the case when the system is 
in contact with the ground and depends on 
the ground stiffness (soft versus hard). 
In fact,  assuming that gravity is fully compensated (this means $K_{L2}=0$), the free leg motion with the torque loop closed is not internally stable (as was shown in the previous section there is a pole-zero 
cancellation on the stability boundary). Further, it is not obvious that maximizing the bandwidth of the 
torque loop is always consistent with the specifications for the outer impedance loop. In the approach 
presented in this paper it was decided to design the controller gains so that the torque loop gain has a 
phase margin larger than $30^\circ$ and a gain margin larger than 12 $dB$ for the upper and lower bounds of 
$J_{L2}$   and  $K_{L2}$ and with a velocity compensation gain $\alpha=0.94$. This would result in a 
satisfactory response. In addition, the closed loop torque response was required to be stable for all 
values of the velocity compensation gain $\alpha$  between zero and one (but the gain and phase margins can 
be less than 12 dB or 30 degrees, respectively).
A set of controller gains satisfying the given 
specifications is found by using the Matlab \textit{SISOtool}.
\begin{equation}
P_t=0.382\beta \quad \quad I_t=18\beta \quad\quad 25>\beta>0
\label{eq:torqueGains}
\end{equation}
Changing the gain  $\beta$ only affects the gain of the controller 
but the controller zero remains 
fixed. This gives ample freedom to investigate the effects 
of increasing the torque loop gain $\beta$  and hence increasing 
the closed loop torque bandwidth when the outer loop specifications are considered.
We must remark that, when defining the torque bandwidth, we consider that the 
system is free to move in the air. Then the bandwidth is the frequency 
where the torque amplitude decreases by -3 $dB$ respect to the reference.

\subsubsection{Harmonic drive torque ripple compensation}
One drawback of using a harmonic drive gearbox 
is that it introduces torque ripples ($T_{fr}$ in \eqref{eq:motorloadDyn}). The
problem is related to the working principle of 
the gearbox that is based on the motion of an
elliptic element (wave generator). 
This motion creates torque fluctuation with a fundamental
frequency which is twice the wave generator angular velocity. 
While this disturbance is normally neglected in position control schemes, because it is passively filtered out by the inertia of
the system, conversely it has a detrimental 
effect on torque control and creates vibrations and
wearing of the components. 
A way to mitigate this problem is to add a lead/lag compensator (notch) in series to the PI controller in order to add enough phase lead at the resonance of the transfer function between  $T_{fr}$ and $T_l$ where the ripple is more prominent, as illustrated in Sec. 3.7, Ch. 4, p. 50 in \cite{focchi13PhDthesis}.

\subsection{Impedance control}
An impedance controller is added as an outer loop 
as shown in Fig. \ref{fig:blockDiagram}. The controller 
gains $P_{gain}$  and $D_{gain}$  represent the stiffness and 
damping for the joint.  The output $T_{l_{ref}}$ of the controller provides the 
reference torque for the inner loop:
\begin{equation}
{T_{l_{ref}}} = {P_{gain}}({\theta _{L1_{ref}}} - {\theta _{L1}}) - {D_{gain}}{\dot \theta _{L1}} 
\label{eq:PDimpedance}
\end{equation}
while $\theta _{L1_{ref}}$ is the desired trajectory for the joint
position. The term involving the link velocity feedback is implemented 
using an averaging filter to reduce the effects of encoder quantization. $T_{ID}$ (see Fig. \ref{fig:blockDiagram}) is an external 
compensation torque (e.g. inverse dynamics) that can be 
added to remove the effects of gravity and inertia (and thus reduce position tracking 
errors):

\begin{equation}
T_{ID} = (J_{L1} + J_{L2})\ddot{\theta}_{L2_{ref}} + mgl_{com} sin(\theta_{L2})
\end{equation}

\noindent where $l_{com}$ is the distance of the leg center of mass from the joint axis. 
For HyQ, a range of values for the impedance loop gains that is considered 
to be sufficient for walking, trotting and running tasks is $P_{gain}\in[1, 2000] Nm/rad$ and 
$D_{gain} \in[1,50]  Nms/rad$.

\section{Stability regions}
\label{sec:simExp}

The analysis here is limited to the abduction-adduction electric 
joint of HyQ with a variable load inertia that depends on the configuration of the leg joints. In 
particular, given a range of impedance parameters, $P_{gain}$  and $D_{gain}$, this analysis will 
assess how the region of closed loop stability is affected by varying the torque controller gain 
($\beta$), the velocity compensation gain ($\alpha$), the number of samples 
used in the averaging filter ($N_{av}$) and the sampling time ($T_s$).

\begin{figure*}[!thp]
\centering  
\begin{subfigure}[b]{0.25\textwidth}
                \centering
                \includegraphics[width=\columnwidth]{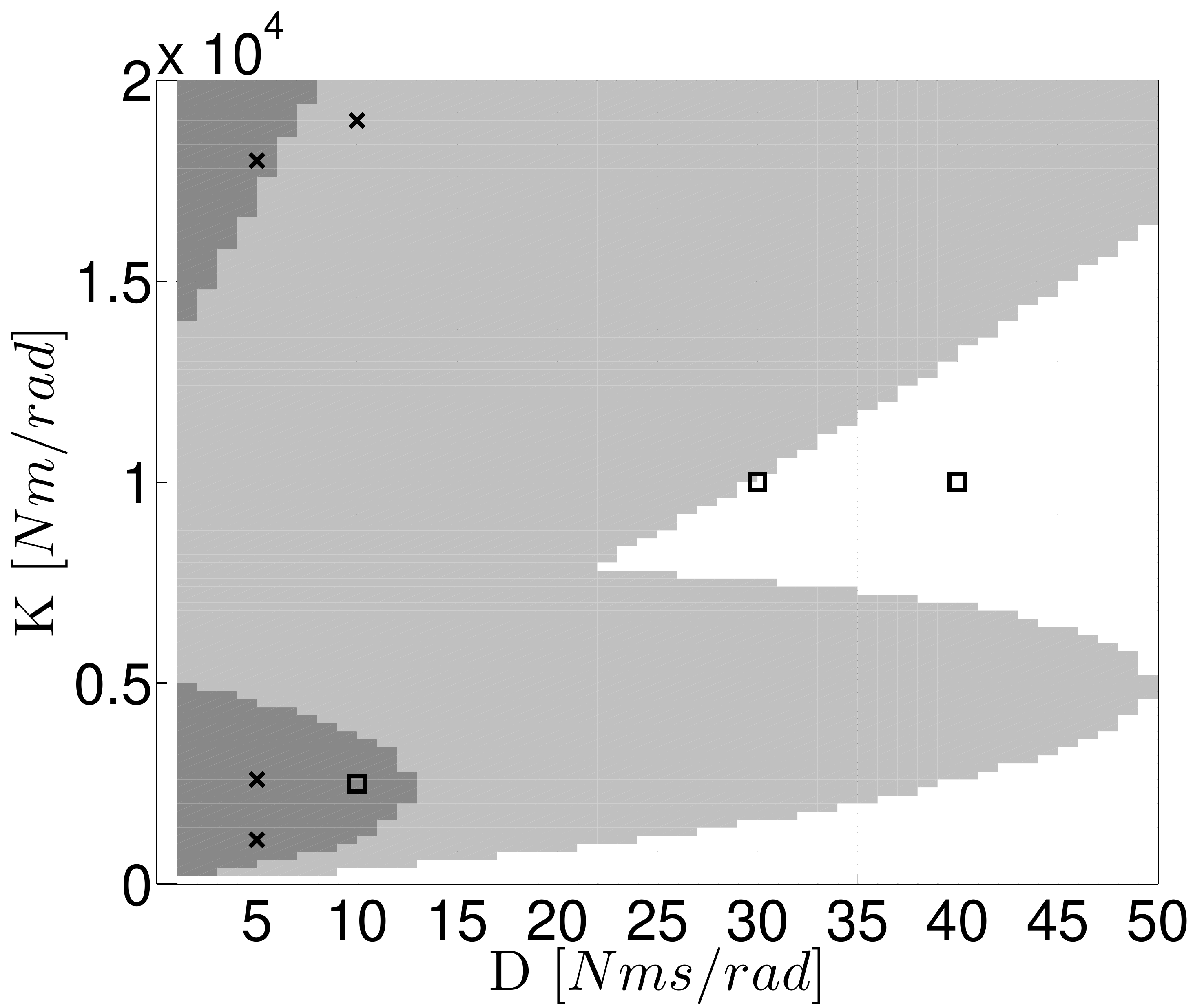}
                \figurecaption{$\beta=1$}
                \end{subfigure}     
                \hspace{-0.2 cm}
\begin{subfigure}[b]{0.25\textwidth}
                \centering
                 \includegraphics[width=\columnwidth]{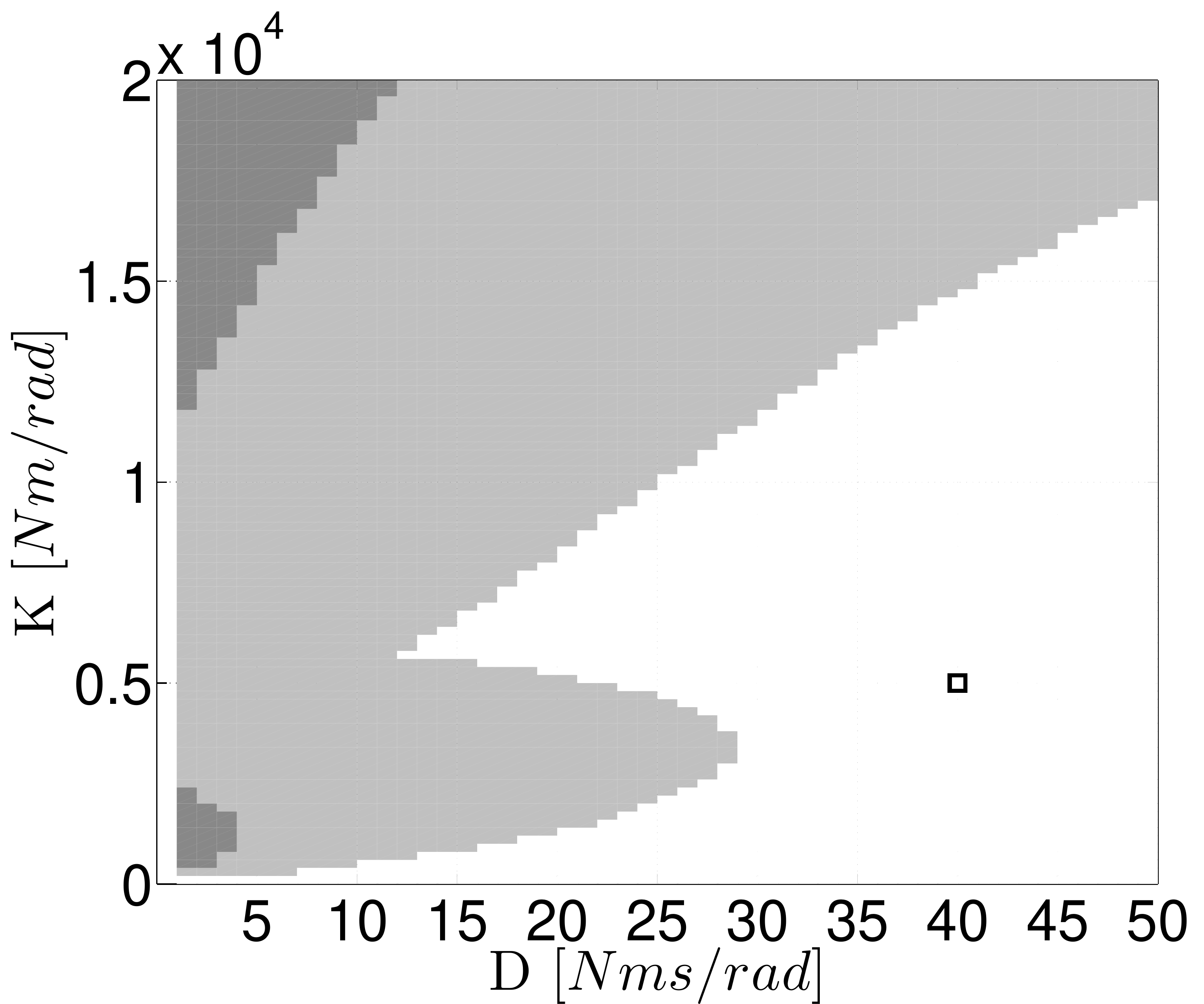}
                 \figurecaption{$\beta=2$}
                 \end{subfigure}
                 \hspace{-0.2 cm}
\begin{subfigure}[b]{0.25\textwidth}
                \centering
                 \includegraphics[width=\columnwidth]{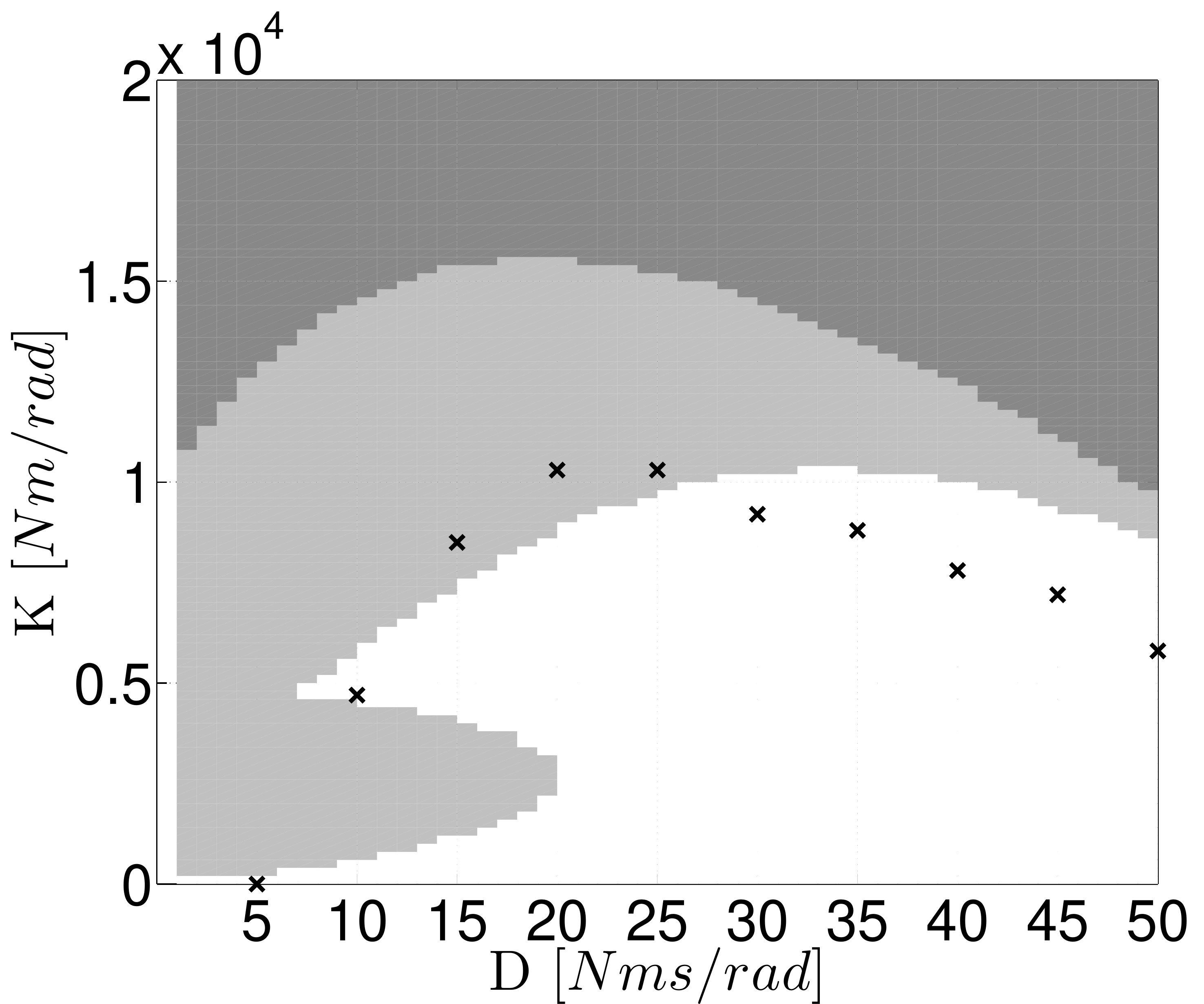}
                 \figurecaption{$\beta=4$}
                 \end{subfigure}
                 \hspace{-0.2 cm}
\begin{subfigure}[b]{0.25\textwidth}
          		 \centering
                 \includegraphics[width=\columnwidth]{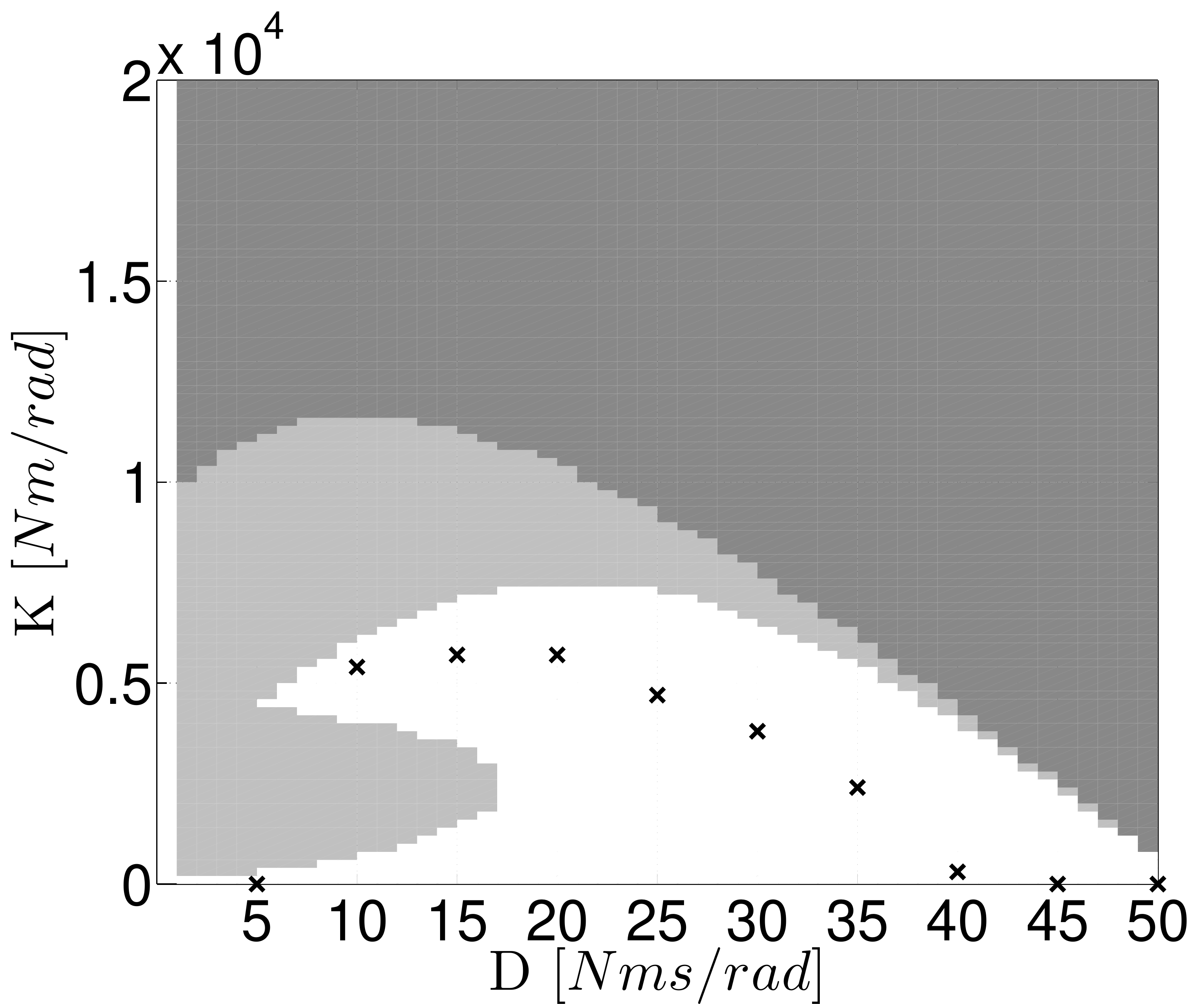}
                 \figurecaption{$\beta=6$}
                 \end{subfigure}
                 \hspace{-0.2 cm}
\figurecaption{Stability regions varying the torque controller gain $\beta$ 
(with $\alpha=0.94$, $N_{av}=4$ and $T_s=1ms$).
White area corresponds to the stable region; light grey is a stable region with 
a phase margin of less than 30 degrees and the dark area is the 
unstable region. Crosses and squares denote unstable and stable experimental points 
respectively.}
\label{fig:beta_region}
\end{figure*}

\begin{figure*}[!thp]
\centering  
\begin{subfigure}[b]{0.25\textwidth}
                \centering
                 \includegraphics[width=\columnwidth]{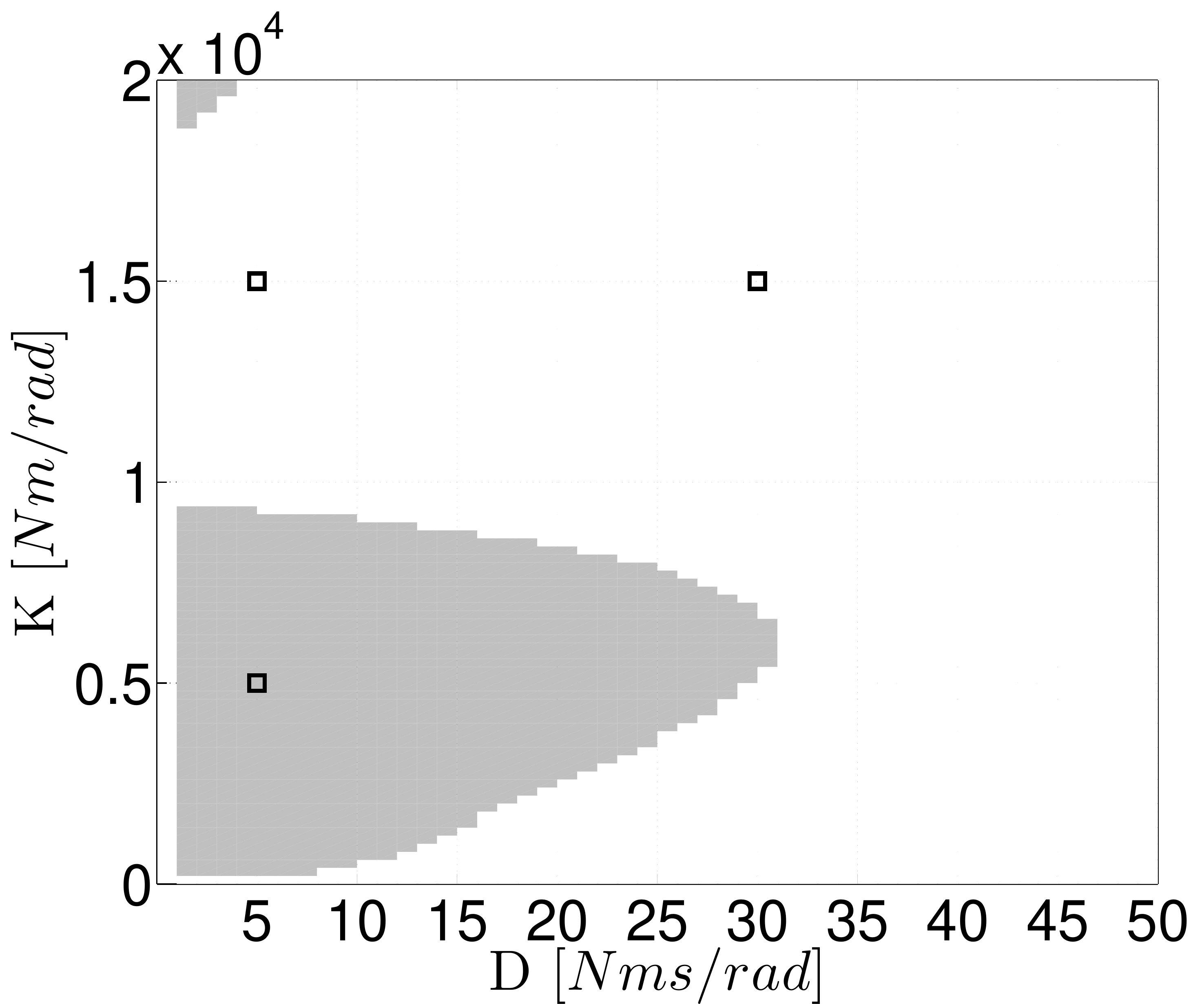}
                 \figurecaption{$\alpha=0$}
                 \end{subfigure}
                 \hspace{-0.2 cm}
\begin{subfigure}[b]{0.25\textwidth}
                \centering
                 \includegraphics[width=\columnwidth]{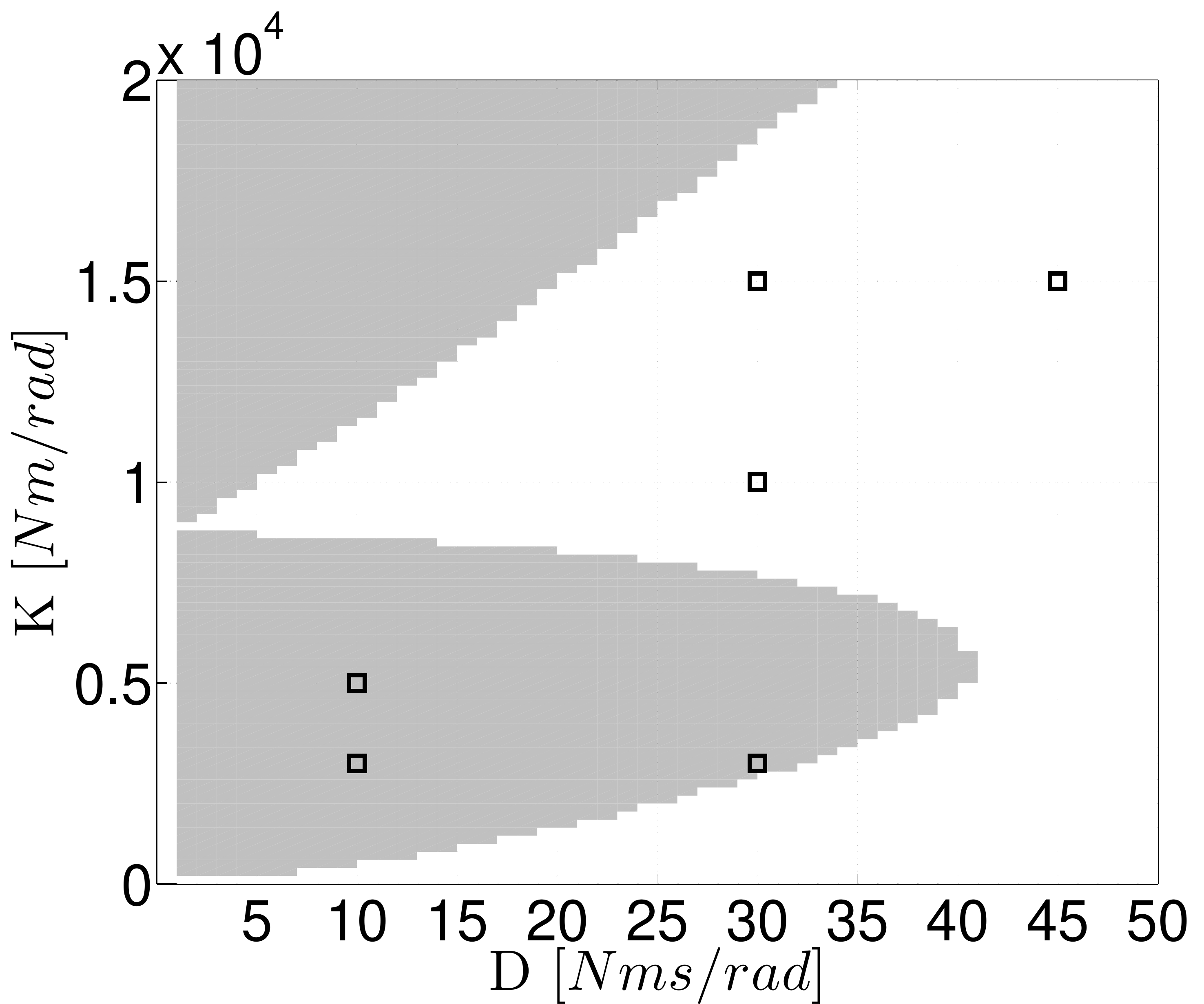}
                 \figurecaption{$\alpha=0.5$}
                 \end{subfigure}
                 \hspace{-0.2 cm}
\begin{subfigure}[b]{0.25\textwidth}
          		 \centering
                 \includegraphics[width=\columnwidth]{matlab/30Pm/PItorque1.pdf}
                 \figurecaption{$\alpha=0.94$}
                 \end{subfigure}
                 \hspace{-0.2 cm}
\begin{subfigure}[b]{0.25\textwidth}
                \centering
                \includegraphics[width=\columnwidth]{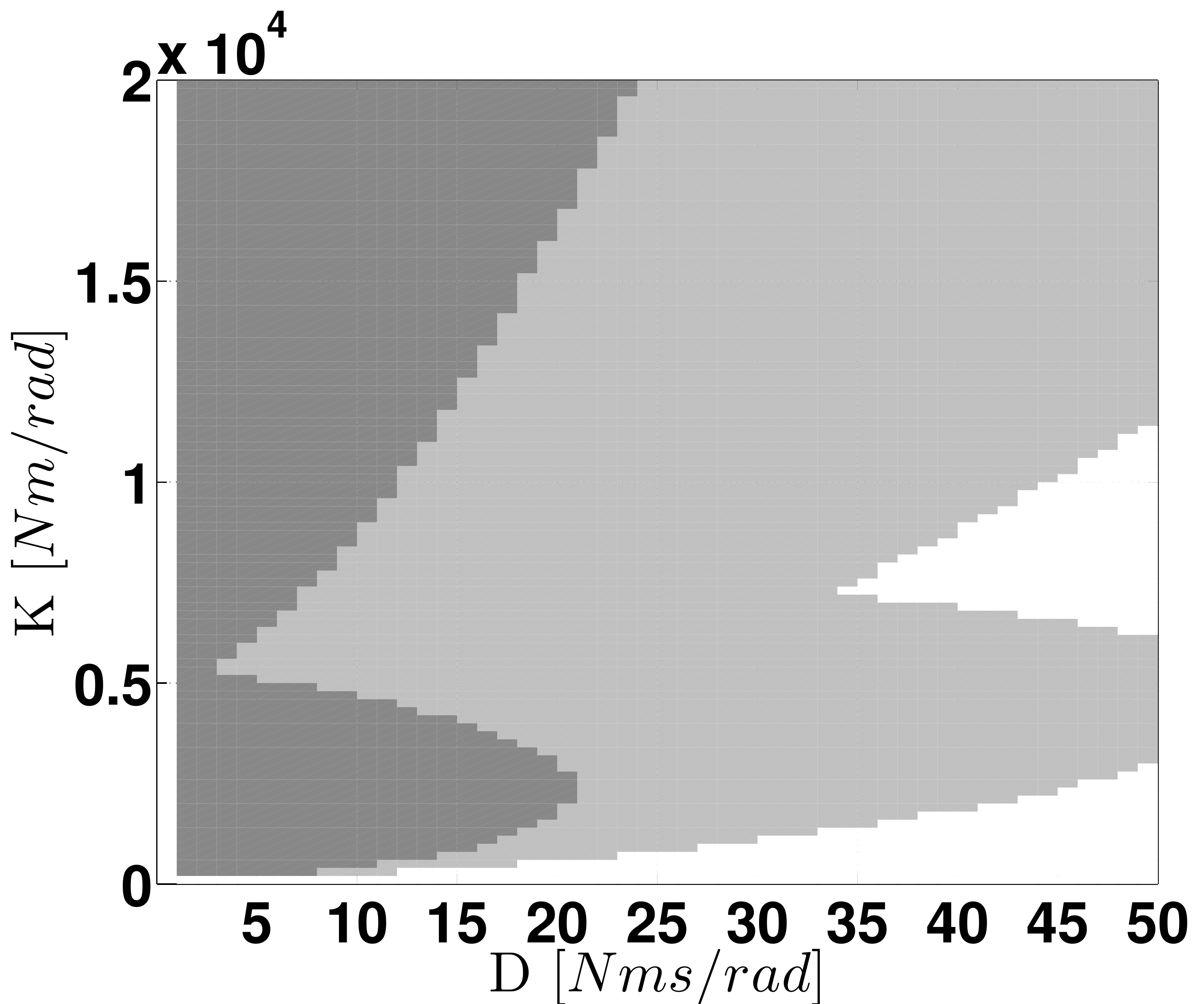}
                \figurecaption{$\alpha=1.2$}
                \end{subfigure}
                \hspace{-0.2 cm}     
\figurecaption{Stability regions varying the velocity compensation 
gain $\alpha$ (with $\beta=1$, $N_{av}=4$ and $T_s=1ms$). 
White area corresponds to the stable region; light grey is a stable region with 
a phase margin of less than 30 degrees and the dark area is the 
unstable region. Crosses and squares denote unstable and stable experimental points 
respectively. }
\label{fig:alpha_region}
\end{figure*}

\begin{figure*}[!thp]
\centering  
\begin{subfigure}[b]{0.25\textwidth}
                \centering
                 \includegraphics[width=\columnwidth]{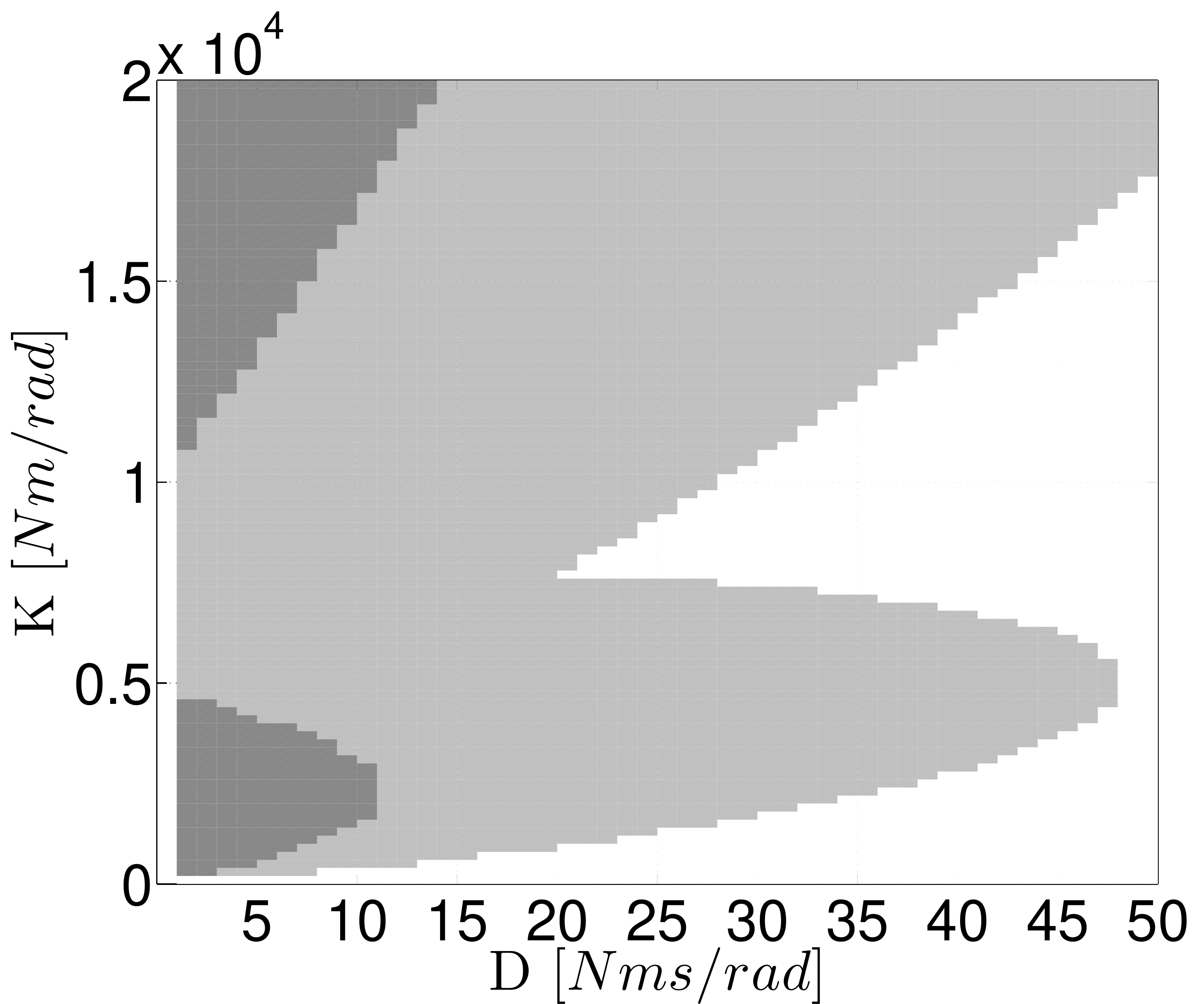}
                 \figurecaption{$N_{av}=1$}
                 \end{subfigure}
                 \hspace{-0.2 cm}
\begin{subfigure}[b]{0.25\textwidth}
                \centering
                 \includegraphics[width=\columnwidth]{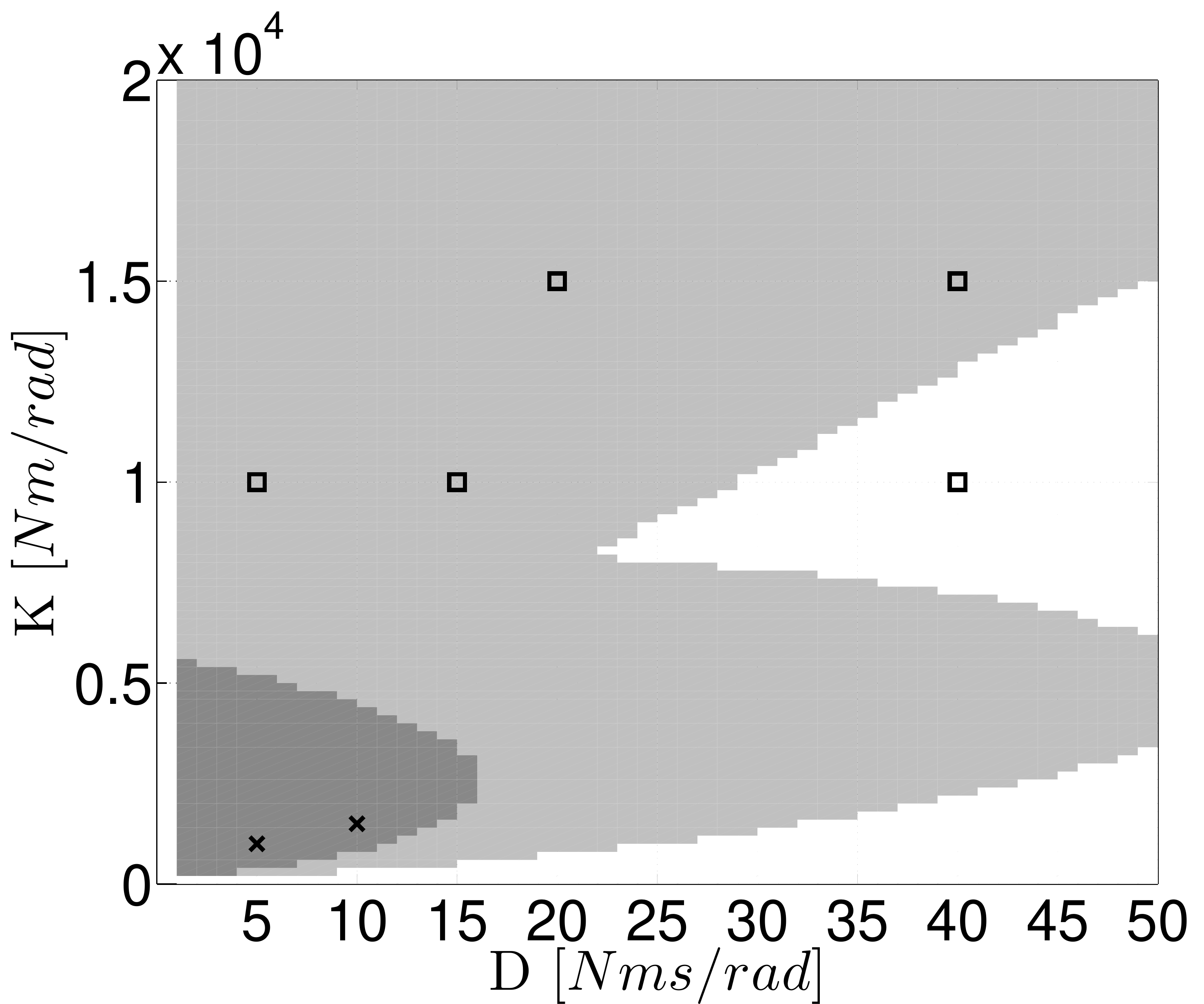}
                 \figurecaption{$N_{av}=10$}
                 \end{subfigure}
                 \hspace{-0.2 cm}
\begin{subfigure}[b]{0.25\textwidth}
          		 \centering
                 \includegraphics[width=\columnwidth]{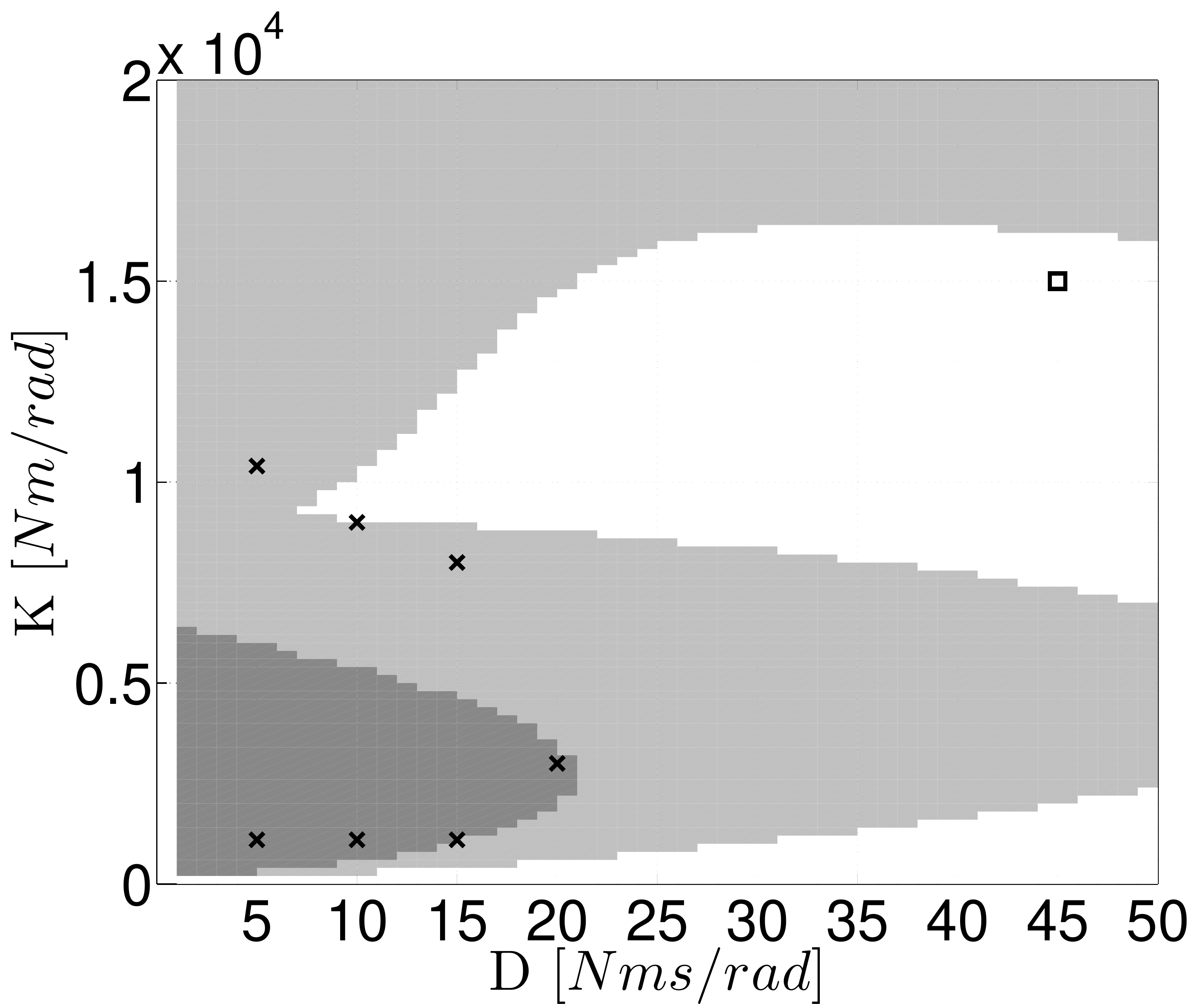}
                 \figurecaption{$N_{av}=20$}
                 \end{subfigure}
                 \hspace{-0.2 cm}
\begin{subfigure}[b]{0.25\textwidth}
                \centering
                \includegraphics[width=\columnwidth]{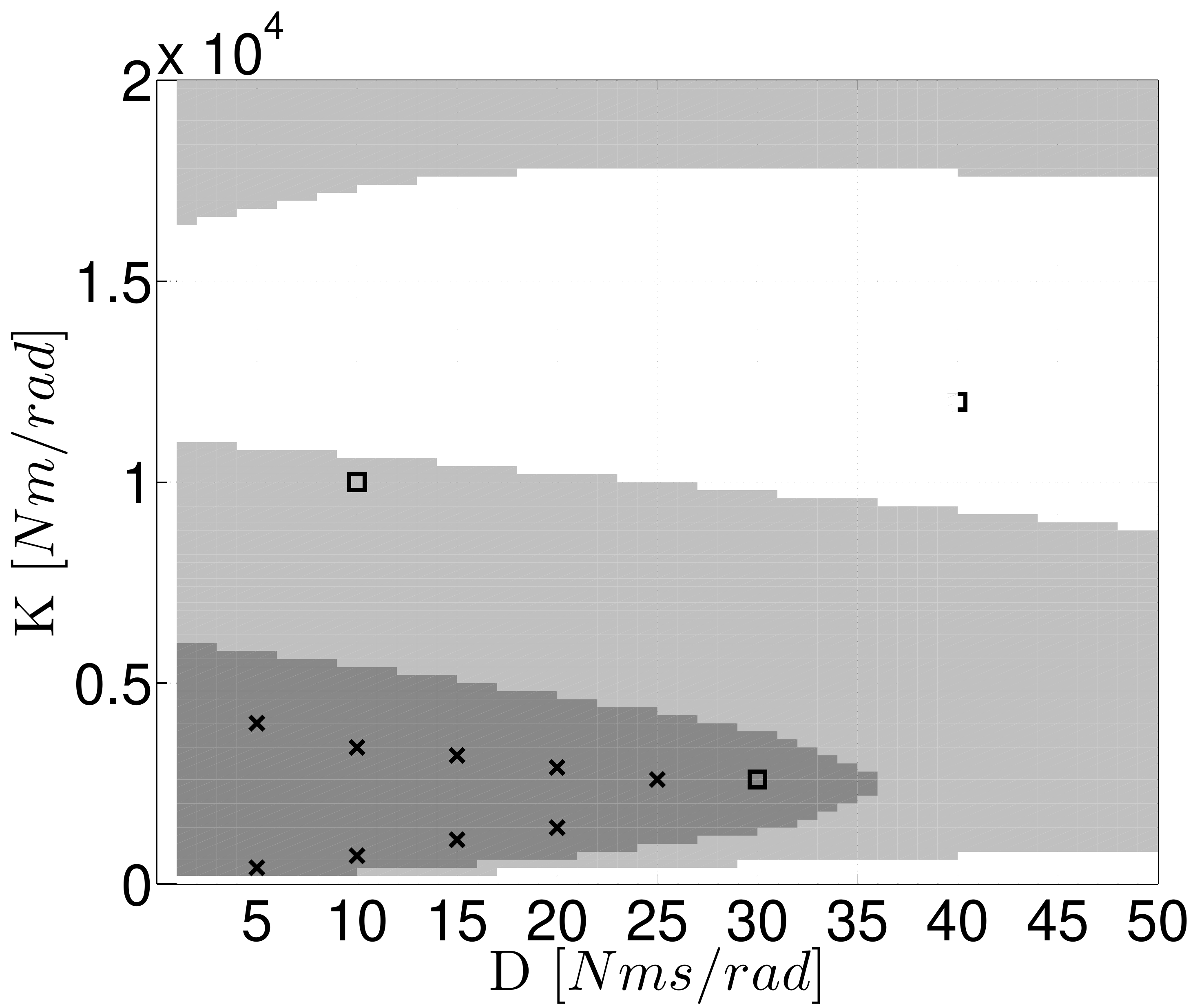}
                \figurecaption{$N_{av}=50$}
                \end{subfigure}    
                \hspace{-0.2 cm}
\figurecaption{Stability regions varying number of samples $N_{av}$ of the 
link velocity filter (with $\beta=1$,  $\alpha=0.94$ and $T_s=1ms$). 
White area corresponds to the stable region; light grey is a stable region with 
a phase margin of less than 30 degrees and the dark area is the 
unstable region. Crosses and squares denote unstable and stable experimental points 
respectively.}
\label{fig:filter_region}
\end{figure*}

\begin{figure*}[!thp]
\centering  
\begin{subfigure}[b]{0.25\textwidth}
                \centering
                 \includegraphics[width=\columnwidth]{matlab/30Pm/PItorque1.pdf}
                 \figurecaption{$T_s=1ms$}
                 \end{subfigure}
                 \hspace{-0.2 cm}
\begin{subfigure}[b]{0.25\textwidth}
                \centering
                 \includegraphics[width=\columnwidth]{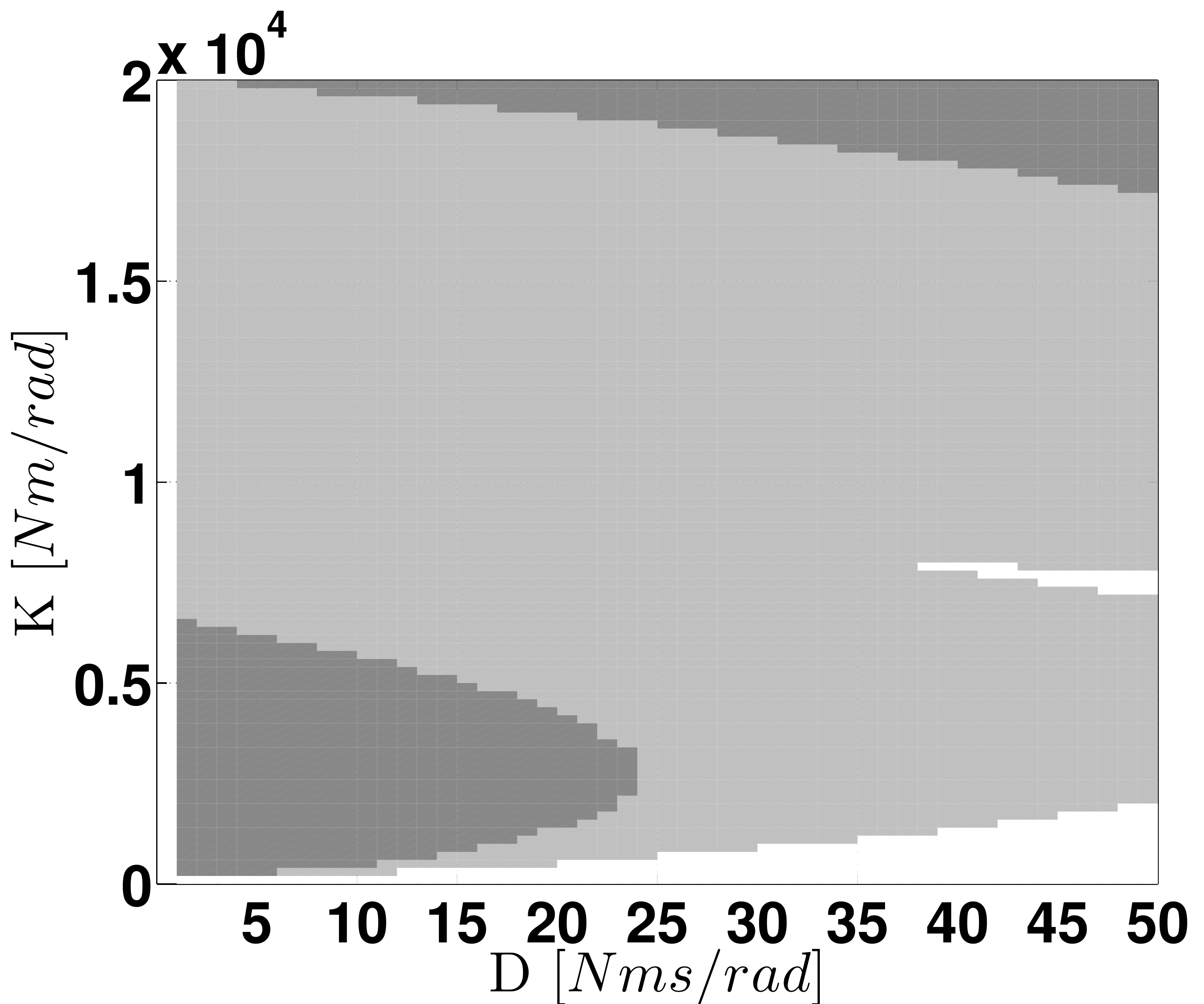}
                 \figurecaption{$T_s=6ms$}
                 \end{subfigure}
                 \hspace{-0.2 cm}
\begin{subfigure}[b]{0.25\textwidth}
          		 \centering
                 \includegraphics[width=\columnwidth]{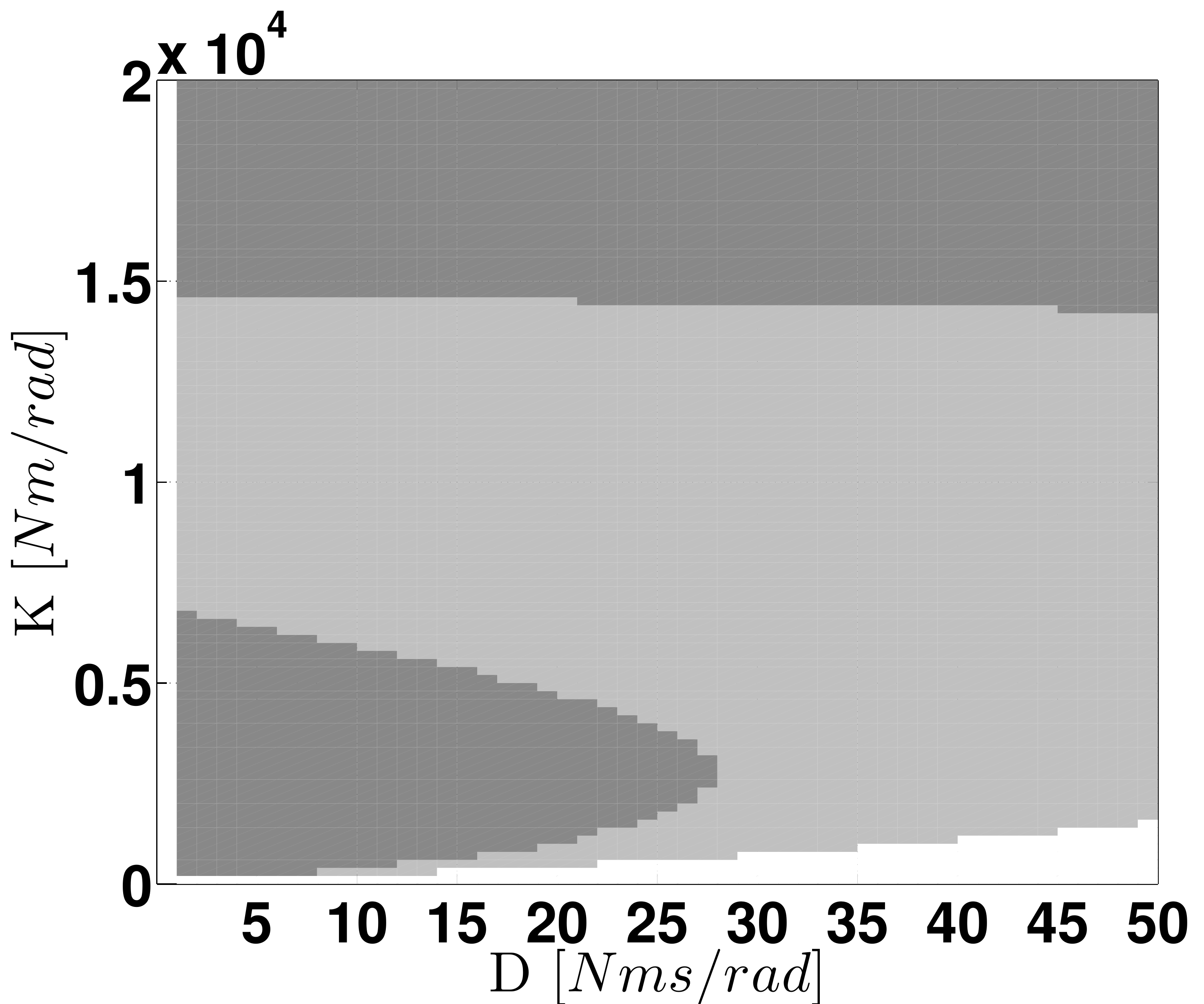}
                 \figurecaption{$T_s=8ms$}
                 \end{subfigure}
                 \hspace{-0.2 cm}
\begin{subfigure}[b]{0.25\textwidth}
                \centering
                \includegraphics[width=\columnwidth]{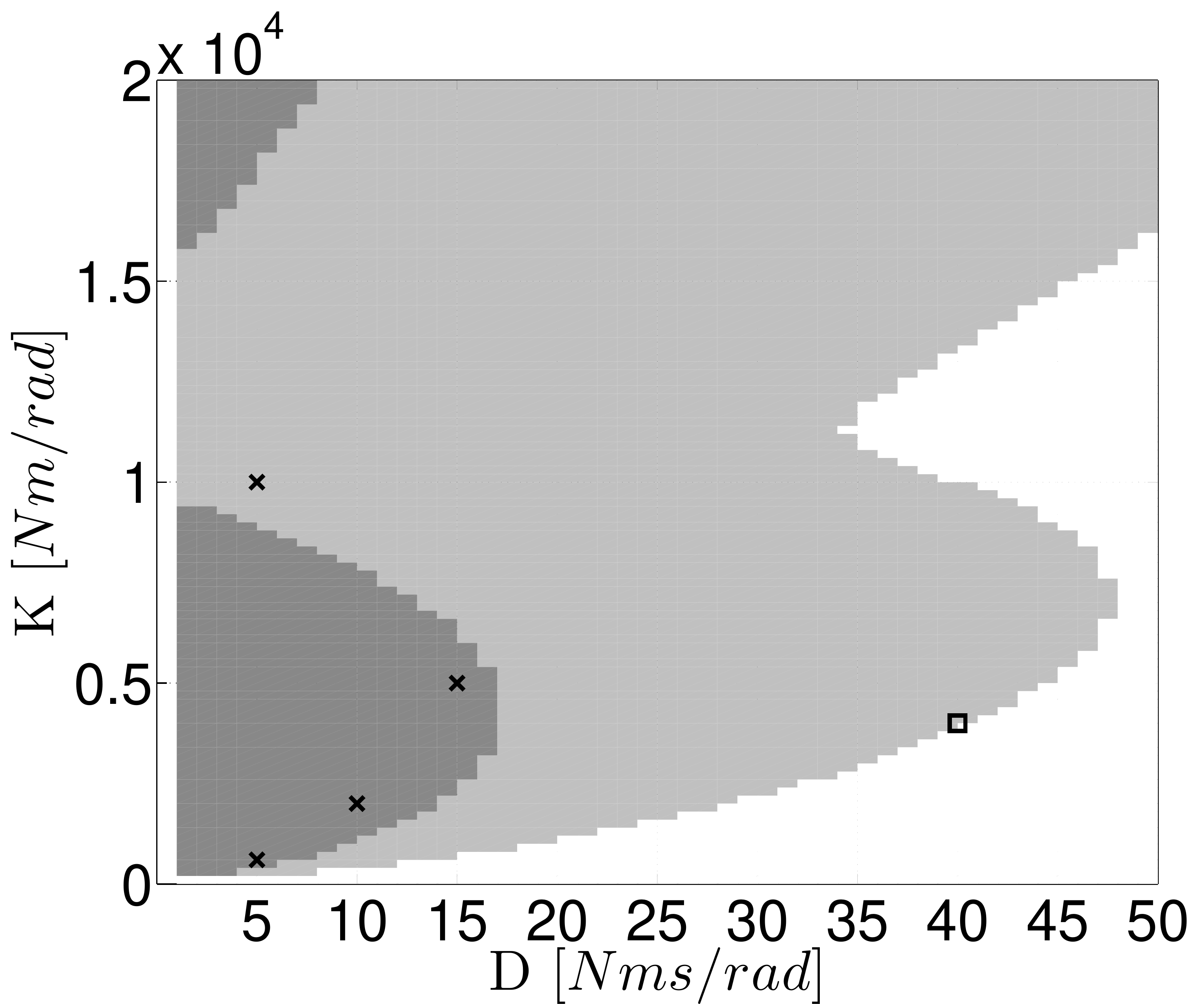}
                \figurecaption{$J_{L2}=0.129$ $kgm^2$}
                \end{subfigure}%
                \hspace{-0.2 cm}    
\figurecaption{(a),(b),(c) Stability regions varying the sampling times $T_s$ (with  
$\beta=1$, $N_{av}=4$, $\alpha=0.94$). (d) Stability region for retracted leg (lower inertia) 
$J_{L2}=0.129$ $kgm^2$ (with $\beta=1$, $\alpha=0.94$, $N_{av}=4$ and $T_s=1ms$). 
White area corresponds to the stable region; light grey is a stable region with 
a phase margin of less than 30 degrees and the dark area is the 
unstable region. Crosses and squares denote unstable and stable experimental points 
respectively.}
\label{fig:sampling_region}
\end{figure*}

The analysis has been performed by varying the stiffness $P_{gain}$  between 1 and 20000 Nm/rad 
and the damping $D_{gain}$  between 1 and 50. 
The upper-bound for the stiffness  was chosen 
such that we could determine the boundary for the stability and passivity regions and is way beyond 
the maximum value used in locomotion. 
The stability of the overall system is 
determined by computing the closed loop eigenvalues and checking that they are inside the 
unit circle. In addition, when closed loop stability is attained, the region where the phase 
margin is less than 30 degrees can be determined. These calculations were carried out in 
\textit{Matlab} using the mathematical model presented in Section \ref{sec:model}. 
The results are 
displayed in Figs. \ref{fig:beta_region}, \ref{fig:alpha_region},  \ref{fig:filter_region} 
and \ref{fig:sampling_region}  where the white area corresponds to the stable region; light 
grey is a stable region with a phase margin of less than 30 degrees and the dark area is the 
unstable region. In the analysis all the regions have been computed for the leg in stretched 
configuration  $J_{L2}=0.439$ $kgm^2$ unless it is otherwise stated.
Fig. \ref{fig:beta_region} shows that as the torque controller gain increases, the 
unstable region for low stiffness and damping decreases but the unstable region for large 
stiffness and/or large damping increases. This clearly illustrates that 
increasing the torque loop bandwidth may not be consistent with the (stability) 
requirements for the outer impedance loop.  
This can be explained if we consider that for any given system 
and controller architecture, there is a limit on 
the maximum loop gain thatcan be achieved, beyond which stability is not ensured
and performance degrades. In a nested architecture both loops contribute to the  loop
gain. If the loop  gain contribution from the torque loop increases (e.g. 
the gain $\beta$ increases), then the contribution from the impedance
loop should reduce otherwise closed loop stability would be lost.
From Figs. \ref{fig:alpha_region} (a), (b) and 
(c) it is clear that increasing the velocity compensation gain
results in an increasing instability region.
Once more this gives further evidence that as the torque loop 
bandwidth increases, the outer loop may become unstable. 
The effects of increasing the number of samples ($N_{av}$) in the velocity averaging filter are shown in Fig. 
\ref{fig:filter_region}. Averaging a large number of samples enlarges the instability region 
for low stiffness values but the unstable region for high stiffness and low damping decreases 
in size.
Figure \ref{fig:sampling_region} clearly shows that for a large sampling time interval the region 
for instability is the largest both for low and high damping and stiffness. The region 
where the phase margin is less than 30 degrees also 
increases as the sampling time increases.
For low leg inertia configuration in Fig. \ref{fig:sampling_region} (d) the instability region for low damping and low stiffness 
increases but the unstable region decreases for low damping and large stiffness. 
The region with a phase margin smaller  than 30 degrees is also larger for the 
low inertia configuration. 

To determine the stability regions experimentally is not an easy task because it would involve  
a vey large number of experiments. Even only finding the stability boundaries is not a simple 
task. Several 
experimental tests were carried out to validate the analytical results predicted by the 
model. The adopted methodology was to start from values $P_{gain}$  and  $D_{gain}$ well inside 
the stable region 
and change the parameters in small steps until the instability was triggered. This enabled us to 
get a rough idea of the boundary of the stability region. The experimental results are displayed in Fig. 
\ref{fig:beta_region}, \ref{fig:alpha_region},  \ref{fig:filter_region} and 
\ref{fig:sampling_region}, where crosses and squares denote unstable and stable points 
respectively.
With the exception of Fig. \ref{fig:beta_region}(c) and \ref{fig:beta_region} (d), overall the 
experimental tests are consistent with the analytical calculations. Fig. \ref{fig:beta_region}
(b), \ref{fig:alpha_region} (a), \ref{fig:alpha_region} (b) and \ref{fig:filter_region} (b) are fully 
in agreement with the theoretical results. For test points near the stability boundaries 
inconsistencies are present suggesting that the model lacks accuracy from a quantitative point of 
view but qualitatively it is correct. The experimental results shown in Fig. 
\ref{fig:beta_region} (c) and \ref{fig:beta_region} (d) are quite different from the expected 
outcome. 
%
%
Even though a precise explanation is currently not available, the authors have several possible 
hypothesis for these discrepancies:  the experimental set-up might have reached some limiting 
conditions, which invalidate the linear analysis (for  example by generating motor voltages that 
exceed the capabilities of the motor drive electronics, or by generating reference torques that 
exceed the range of the torque sensor). Their investigation is part of future work.
\section{Passivity analysis}
\label{sec:passivity}
Another question of interest is whether the closed loop system remains stable when it interacts with a 
passive environment. 
It is well-known that a strictly passive system,
connected to any passive environment, is necessarily stable
\cite{colgate89}. Thus, since most terrain surfaces are passive,
to ensure a stable contact with the environment also the robot joints have
to be passive. This section therefore analyzes the main
factors that influence passivity.
%
Stating that a system is passive, is equivalent to saying that the system is intrinsically dissipative.
This is not always the case when compliance is obtained actively, where the compliant behavior
is emulated by  controlled actuators. In this case, the controller gains can destroy passivity. The requirement to ensure this type of stability for the robot interacting with the environment is the following: 
the port of interaction between the system and the environment, i.e. the driving port impedance, has to be 
passive. For linear time invariant systems this is a necessary and sufficient condition, but it is only a sufficient condition for nonlinear systems.
Let $Z(s)$ denote the driving port impedance transfer function. Then $Z(s)$ is passive if and only if it is positive real \cite{anderson06}. In \cite{colgate94journal} and \cite{colgate86} it has been shown that this is equivalent to:
\begin{itemize}
	\item[1.] $Z(s)$ has no poles in $\Re(s)>0$;
	\item[2.] the phase of $Z(s)$ lies between -90 and 90 degrees.
\end{itemize}
For sampled data control systems, Colgate \cite{colgate94journal} has suggested
an approximate method based on computing the corresponding discrete
time transfer function $Z(z)$, assuming that the port of
interaction is also sampled.
The phase of $Z(z)$ is computed and corrected
by subtracting $\omega T_s/2$ $rad$, where $T_s$
is the sampling time interval. 
Although many studies have been carried out for analyzing the
passivity of sampled-data systems \cite{Colgate1997}, there is still a lack of
information about the influence of the closed-loop torque
control bandwidth on the combinations of stiffness and damping that can be passively
rendered (in the field of haptics called \textit{Z-width}) \cite{colgate94journal}. 
Therefore, this section will show 
that the torque loop performance
plays an important role in determining the range of passively achievable
impedance values.
The discrete time transfer function  $Z(z)$ (impedance) has been computed from the 
link velocity $\dot{\theta}_{L1}$ to the load disturbance $T_{dist}$.  
The transfer function has been computed for two cases: one for the torque 
loop (with velocity compensation) and one for the system after closing the outer impedance loop. 
Then the above-mentioned phase correction is applied to include the fact that the system is sampled.
In the approach presented in this paper the  analysis of passivity has been done by 
varying several parameters to have a better
 understanding of their influence: the gain of the PI torque controller 
 (by varying the gain $\beta$ in Eq. (\ref{eq:torqueGains})), the velocity 
 compensation gain $\alpha$, the sampling time $T_s$, and the number of 
 samples  $N_{av}$ of the averaging filter of the link velocity.
For each set of parameters the analysis was  performed by first checking the 
stability of $Z(z)$ and then 
verifying that the corrected phase of $Z(z)$ was in the range $\pm90^\circ$ 
for frequencies up to the 
Nyquist frequency. If these conditions are not satisfied then $Z(z)$ is not 
passive for the particular 
set of parameter values. The results of this analysis are summarized in 
Table \ref{tab:passivity}.
\begin{center}
\centering
\vspace{-0.2cm}
\captionof{table}{Passivity}
\label{tab:passivity}
\renewcommand{\arraystretch}{1.0}
\begin{tabular}{ p{2.5cm} ccc}
\hline \hline
\multirow{3}{*}{}& Torque      &	Imp. loop	 &	Imp. loop	\\
				 & loop                         & 			$P_{gain}=200$   & $P=20000$        \\
				 &							   &     		$D_{gain}=10$   &  $D=50$		    \\
\hline
$\beta=1$ &	     No						& Yes	 					 & Yes						\\
\hline
$\beta=0.5$&	     No						& Yes	 					 & No						\\  
\hline
$\beta=2$&	     No						& Yes	 					 & No						\\  
\hline
$\beta=4$&	     No						& Yes	 					 & Unstable			\\  \hline
$\beta=6$&	     No						& Yes	 					 & Unstable			\\  
\hline
$\alpha=0$ 			  &	     Yes					& Yes	 					 & Yes						\\
\hline
$\alpha=0.5$           &	     Yes					& Yes	 					 & Yes						\\
\hline
$T_s=4\cdot10^{-3} [s]$  &	     No						& Yes	 					 & No						\\
\hline
$T_s=2\cdot10^{-3} [s]$  &	     No						& Yes	 					 & No						\\
\hline						
$T_s=0.5\cdot10^{-3} [s]$&	     No						& Yes	 					 & Yes						\\
\hline
Averag. $N_{av}=1$		  	  &	     No						& Yes	 					 & Yes						\\
\hline
Averag. $N_{av}=10$		  &	     No						& Yes	 					 & Yes						\\
\hline
Averag.$N_{av}=20$		  &	     No						& Yes	 					 & Yes						\\
\hline
Averag. $N_{av}=50$	  	  &	     No						& Yes	 					 & Yes						\\
\hline
Low leg Inertia (ret.) $J_{L2}=0.129$ $kgm^2$	  	  &	\multirow{3}{*}{  No}& \multirow{3}{*}{Yes} & \multirow{3}{*}{Yes}\\	
\hline
\hline
\vspace{-0.2cm}
\end{tabular}
\end{center}
In this table the nominal set of parameters are:  
$\alpha=0.94$, $N_{av}=4$ samples, leg inertia $J_{L2}=0.439$ $kgm^2$, $P_t=0.38$ and $I_t=18$.
The overall (impedance + torque loop) system is always 
passive for low impedances $P_{gain}=200 \quad D_{gain}=10$ while passivity 
might be destroyed  when the torque controller gain $\beta$ (and so the torque bandwidth) 
increases or the sampling frequency decreases. In particular it can be noticed that 
when the torque gain is larger than or equal to 4 the closed loop system with the impedance 
loop becomes unstable. This is a clear indication that increasing the bandwidth of 
the torque loop is not always consistent with the requirements of the position loop.  
When only the torque loop is closed the system is almost never passive except 
for low values of  $\alpha$. The table shows that the velocity compensation is 
the key parameter affecting the  passivity if the torque loop alone is considered. 
In particular when the amount of velocity compensation increases, the inner torque 
loop becomes not passive and therefore the torque control system alone can become unstable 
when the leg interacts with some environments. Further analysis showed that the 
system (without closing the impedance loop) becomes unstable if the leg is in contact with
an environment with a stiffness $K_{L2}$ between 72 $Nm/rad$ and 3500 $Nm/rad$.
This has been verified with experimental tests using the test setup 
depicted in Fig. \ref{fig:passive_test_setup} by commanding a leg motion to 
have an impact against a physical spring. This spring is positioned in order to create a certain stiffness $K_{L2}$.
\begin{center}
  \centering  
  \includegraphics[width=0.7\columnwidth]{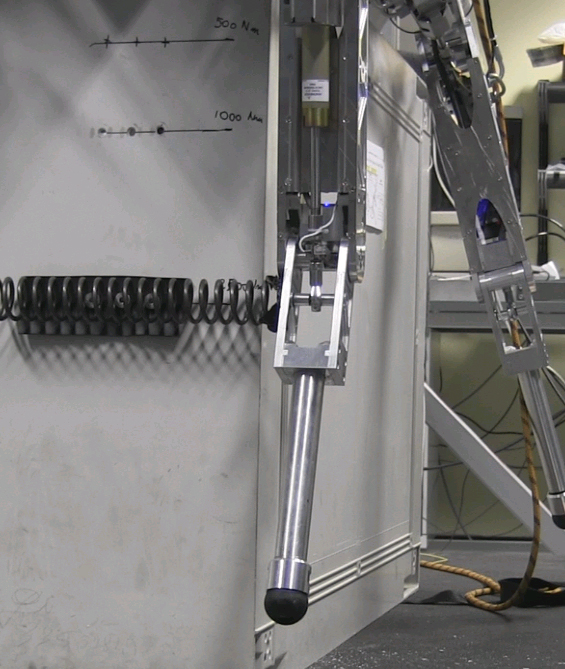}                
  \figurecaption{Experimental setup for passivity tests. The rotational stiffness 
	$K_{L2}$ is  obtained by positioning a linear spring 
	at a certain distance from the $HAA$ axis.}
  \label{fig:passive_test_setup}
\end{center}
Figure \ref{fig:phase_impedance} shows the phase
of $Z(z)$ when only the torque loop is closed ($\beta=1$ and $\alpha=0.94$) 
and when both the torque and the impedance loops are closed
($P_{gain}=200$, $D_{gain}=10$). 
The curves illustrate that, in the case 
of the torque loop alone, the phase of $Z(z)$ exceeds 90 degrees  
between 10 and 50 $rad/s$, indicating the loss of passivity, while, 
when the outer loop is closed, the phase always remains within $\pm 90$ degrees,
demonstrating that the passivity property is ensured 
as indicated in Table \ref{tab:passivity}.
\begin{center}
  \centering
	\includegraphics[width=1.0 \columnwidth]{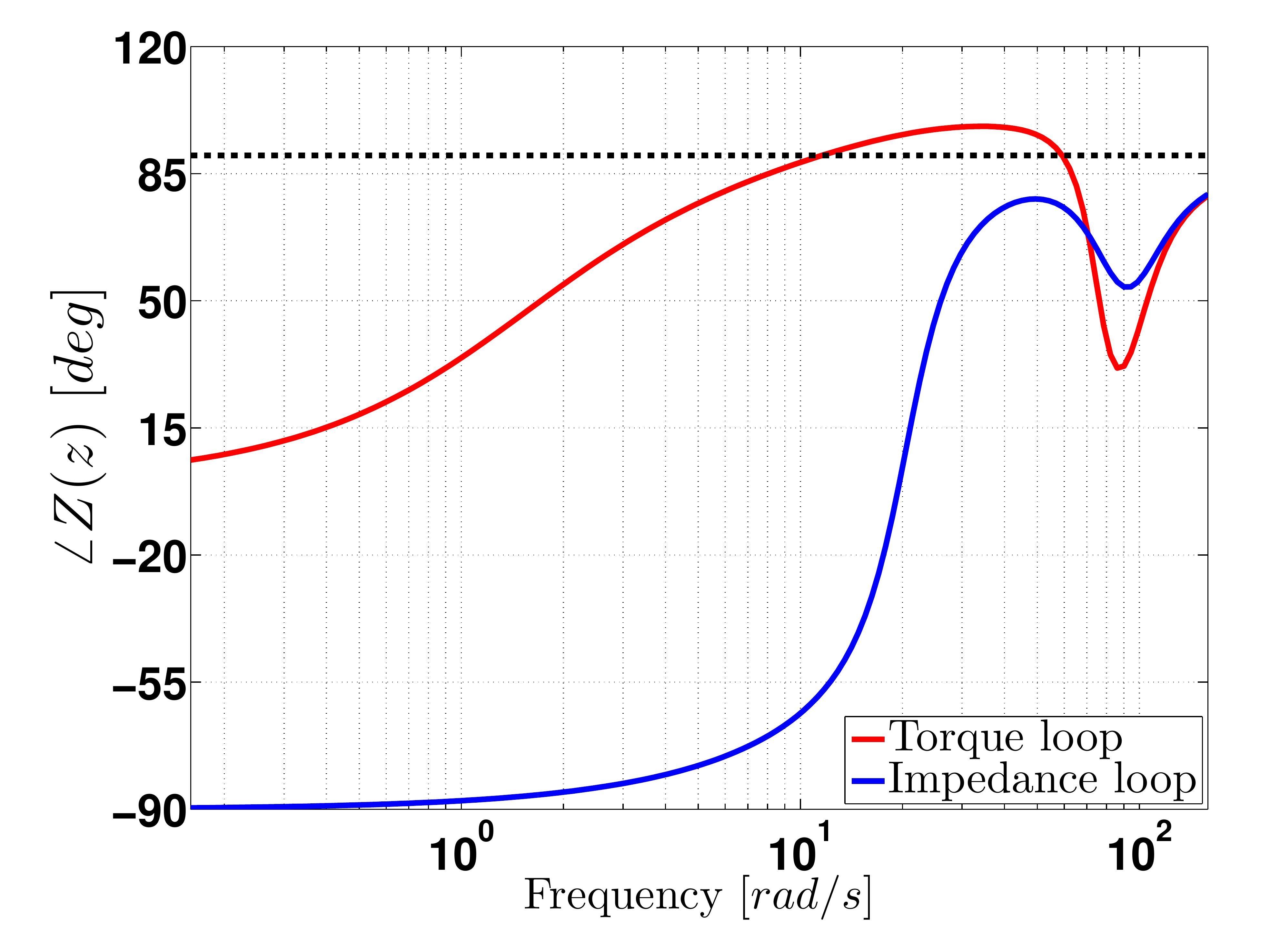}
	\figurecaption{Simulation. Phase plot of the driving port impedance 
	$Z(z)$ when only the torque loop is closed ($\beta=1$ and $\alpha=0.94$) (red plot)
	and when both the torque and the impedance loops are closed 
	($P_{gain}=200$, $D_{gain}=10$) (blue plot). The black line shows the limit of 90 degrees.}
	\label{fig:phase_impedance}
\end{center}
\section{Conclusions and future work}
\label{sec:conclusions}
This paper presented a methodology for designing joint impedance 
controllers based on an inner torque loop and a positive velocity feedback 
loop. In particular, it was shown that the positive velocity feedback can be 
used to increase the closed loop bandwidth of the torque loop without the need 
of a complex controller. 
It has been demonstrated that besides the sampling 
frequency and filtering, the bandwidth of the torque loop has a strong 
influence on the range of impedance parameters ($Z$-width) that exhibit a passive 
and/or stable behavior. 
Indeed, larger inner loop bandwidth can be beneficial for disturbance rejection and 
to improve the tracking
of the impedance (enlarge the range of frequencies for which the desired impedance 
is emulated by the system) but, at the same time, it can reduce the region of stable impedance 
parameters. 
This fact can limit the performance and 
versatility of a robot.
Thus, the highest possible bandwidth for the torque loop might not 
be the best solution for all situations. 
It is therefore important to find a balance between 
the torque loop requirements (e.g. to have a good torque tracking) and the stability/passivity specifications of the 
overall system. Furthermore, it has become evident that, even for simple controllers, 
the design problem is challenging and that there are competing trade-offs to consider
when selecting the controller gains. 

This suggests that there is a need for a 
problem formulation that can encompass the design objectives in a more 
systematic and generic framework. There are a number of areas that need 
further research and are left for future work. 
Adaptive schemes (e.g. gain scheduling) can be 
developed to modify the torque controller gains to satisfy the constraints of stability/passivity 
given by the desired impedance parameters specified by the system designer.
The torque gains can also be  modified depending on the load inertia variations that is changing
with the leg configuration.  
In addition, varying the location of the PI torque controller zero can provide improvements in 
performance. The torque controller architecture can also be enhanced to reduce the effects of 
torque ripples arising in the gear transmission system (drive jitter). Since increasing the torque 
controller gain has been shown to be detrimental, this option for mitigating the drive jitter can 
be discarded and more different solutions must be found. Finally, there is a need to develop strategies to quantify the range of impedances 
that are required for specific tasks. At present there are some guidelines that only provide 
qualitative results, for example a high stiffness ($P_{gain}$) is specified when there is contact 
with a compliant environment and the positioning accuracy is important. On the other hand, a low 
stiffness is used to maintain small contact forces or when the environment is stiff. Similarly, 
large damping values ($D_{gain}$) are needed to reduce vibrations or to quickly dissipate energy.
Future work also encompasses an extension of the methodology presented in this paper to include
multi-degree of freedom systems.

\begin{ack}  
\noindent 
This research has been funded by the Fondazione Istituto Italiano di Tecnologia. J. Buchli is supported by a Swiss National Science Foundation professorship.
\end{ack}


\vspace{.5\baselineskip} {\selectfont\scriptsize

\parpic[l][b]{%
  \begin{minipage}{\dimexpr\linewidth-2.7cm-3.3cm}
    \includegraphics[width=\linewidth]{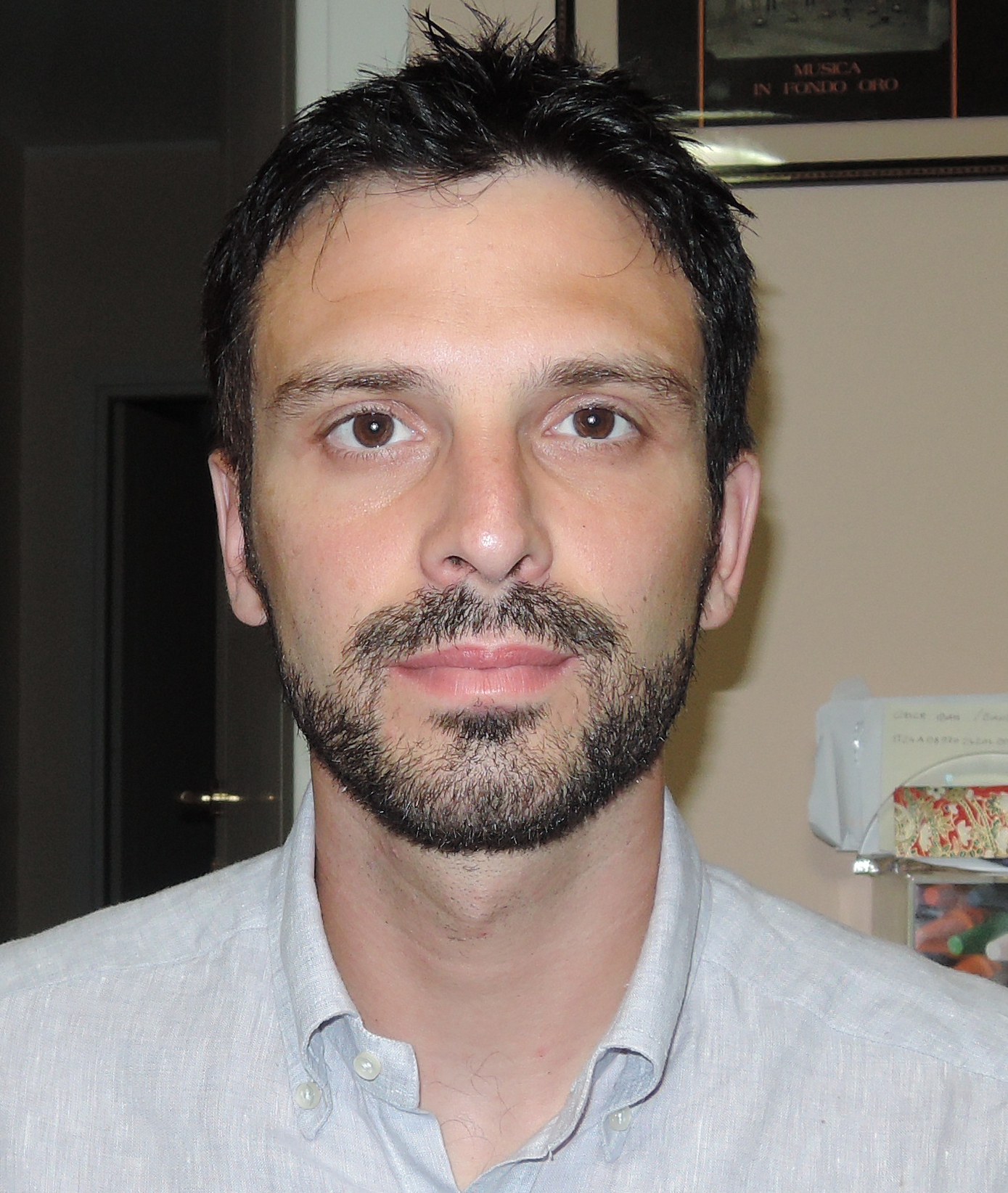}%
  \end{minipage}} 
{\bf Michele Focchi} is currently a Post-Doc researcher at the Advanced Robotics department of Istituto Italiano di 
Tecnologia. He received  both the B.Sc. and the M.Sc. in Control System Engineering from Politecnico di Milano in 
2004 and 2007, respectively. Until 2009 he worked for the 
R$\&$D department of Indesit company. In 2009 he joined the Advanced Robotic Department (ADVR) at Istituto Italiano 
di Tecnologia, (Genova, Italy) developing a prototype of a novel concept of air-pressure driven micro-turbine for 
power generation for which he obtained an international patent and several awards. In 2013 he got a Ph.D. degree in 
robotics, getting involved in the Hydraulically Actuated Quadruped Robot project  where he developing  low-level 
controllers for locomotion purposes. Currently his research interests range from  planning of dynamic motions to 
whole body control.
\par}

\vspace{.5\baselineskip} {\selectfont\scriptsize
\parpic[l][b]{%
  \begin{minipage}{\dimexpr\linewidth-2.7cm-3.3cm}
    \includegraphics[width=\linewidth]{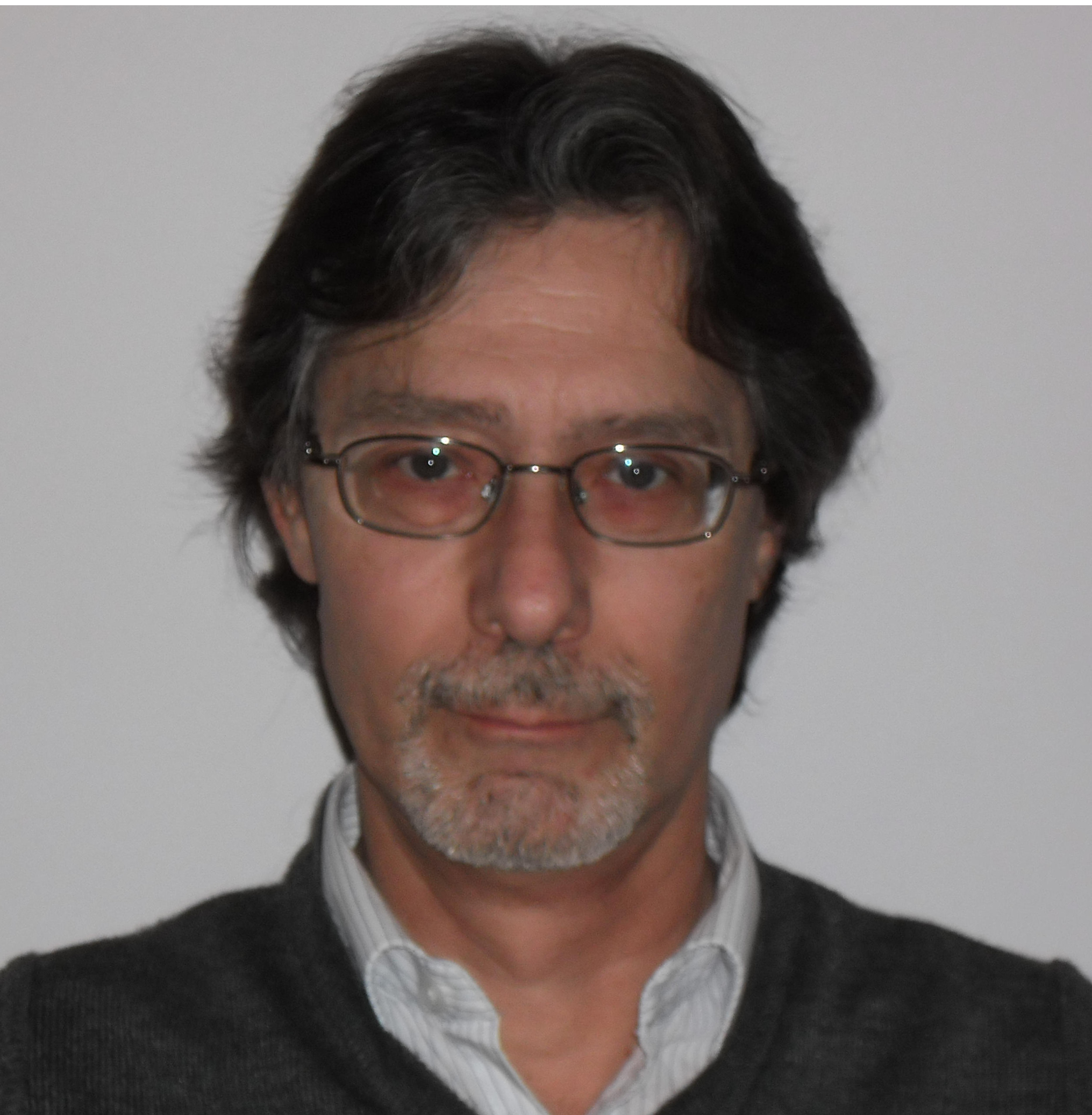}%
  \end{minipage}} 
{\bf G.A. Medrano-Cerda} received the B.Sc. degree in electro-mechanical engineering from the Universidad Nacional 
Autonoma de Mexico in 1977, and the M.Sc. and Ph.D. degrees in control systems from Imperial College, London, in 
1979 and 1982, respectively. From 1982 to 1985 he was an associate professor at the Division de Estudios de
Postgrado, Facultad de Ingenieria, Universidad Nacional Autonoma de Mexico. From 1985 to 1986 he was a research 
fellow at the Department of Engineering, University of Warwick. From 1986 to 2002 he was a lecturer at the 
Department of Electronic and Electrical Engineering at the University of Salford. During this period he set up the 
Advanced control and robot locomotion laboratory at the University of Salford. In 1999 he became a control systems 
consultant at Las Cumbres Observatory (formerly Telescope Technologies Ltd) and later in 2002 he joined the company 
as a senior control engineer pioneering work in H-infinity control system design and implementation for 
astronomical telescopes. Since 2009 he has been a senior research scientist at the Advanced Robotics Department, 
Istituto Italiano di Tecnologia. His research interests are in the areas of robust control, adaptive control, 
modelling and identification, fuzzy systems and advanced robotic applications, in particular to walking robots.
\par}

\vspace{.5\baselineskip} {\selectfont\scriptsize
\parpic[l][b]{%
  \begin{minipage}{\dimexpr\linewidth-2.7cm-3.3cm}
    \includegraphics[width=\linewidth]{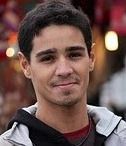}%
  \end{minipage}} 
{\bf Thiago Boaventura}  
       received his B.Sc. and M.Sc. degrees in mechatronic engineering from Federal University of Santa Catarina, Florianopolis, Brazil, in 2009 and the Ph.D. degree in robotics, cognition, and interaction technologies from a partnership between Istituto Italiano di Tecnologia and University of Genoa, Genova, Italy, in 2013. He is a Postdoctoral Researcher with Agile and 
Dexterous Robotics Laboratory, ETH Zurich, Zurich, 
Switzerland. He is mainly involved in the EU FP7 BALANCE project with a focus in the collaborative impedance control of exoskeletons. His research interests include impedance and admittance control, model-based control, legged robotics, optimal and learning control, and wearable robotics.
\par}

\vspace{.5\baselineskip} {\selectfont\scriptsize

\parpic[l][b]{%
  \begin{minipage}{\dimexpr\linewidth-2.7cm-3.3cm}
    \includegraphics[width=\linewidth]{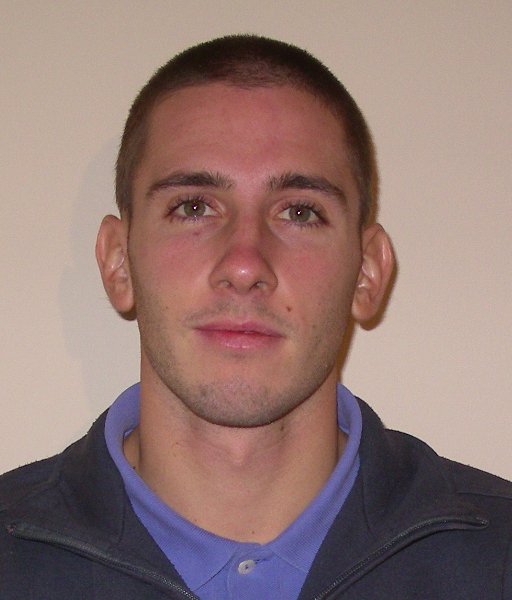}%
  \end{minipage}} 
  {\bf Marco Frigerio} 
received his B.Sc. and M.Sc. degrees in computer science from the University 
of Milano Bicocca, respectively in 2006 and 2008, and the Ph.D. degree in 
Robotics from the Istituto Italiano di Tecnologia (IIT) in 2013. 
He is currently a Post-Doc researcher at the Dynamic Legged Systems lab 
of IIT, where he is involved in the development of hydraulic legged robots. His research interests include software for robotics, software 
architectures, kinematics and dynamics of mechanisms. 
\par}

\vspace{.5\baselineskip} {\selectfont\scriptsize   
\parpic[l][b]{%
  \begin{minipage}{\dimexpr\linewidth-2.7cm-3.3cm}
    \includegraphics[width=\linewidth]{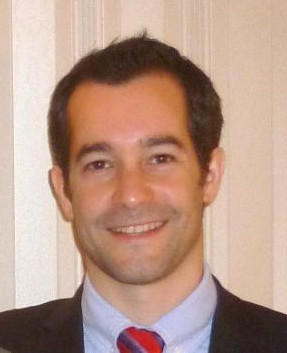}%
  \end{minipage}} 
{\bf Claudio Semini} is the head of the Dynamic Legged Systems lab of the Department of Advanced Robotics at Istituto Italiano di Tecnologia (IIT). He holds an MSc degree from ETH Zurich in electrical engineering and information technology (2005). From 2004 to 2006 he first visited the Hirose Lab at the Tokyo Institute of Technology, followed by work on mobile service robotics at the Toshiba R$\&$D center in Kawasaki, Japan. During his doctorate from 2007-2010 at the IIT he designed and constructed the quadruped robot HyQ and worked on its control. After a post-doc in the same department, in 2012, he became the head of the Dynamic Legged Systems lab. His research focus lies on the construction and control of highly dynamic and versatile legged robots in real-world environments.
\par}

\vspace{.5\baselineskip} {\selectfont\scriptsize 
\parpic[l][b]{%
  \begin{minipage}{\dimexpr\linewidth-2.7cm-3.3cm}
    \includegraphics[width=\linewidth]{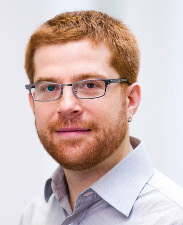}%
  \end{minipage}} 
{\bf Jonas Buchli} received the Diploma degree in electrical engineering from ETH Zurich, Zurich, Switzerland, in 2003 and the Ph.D. degree from Ecole Polytechnique Federale de Lausanne, Lausanne, Switzerland, in 2007. 
From 2007 to 2010, he was a Postdoctoral Researcher with the Computational Learning and Motor Control Laboratory, University of Southern California (USC), where he was the Team Leader of the USC Team for the DARPA Learning Locomotion 
challenge. From 2010 to 2012, he was a Team Leader 
with the Advanced Robotics Department, Istituto Italiano di Tecnologia, Genova, Italy. He is currently an Assistant Professor with the Institute of Robotics 
and Intelligent Systems, ETH Zurich, Zurich, Switzerland, and the Director of 
the Agile and Dexterous Robotics Lab. Dr. Buchli has received a Prospective and an Advanced Researcher Fellowship from the Swiss National Science Foundation (SNF). In 2012, he received a Professorship Award from the SNF.
\par}

\vspace{.5\baselineskip} {\selectfont\scriptsize 
\parpic[l][b]{%
  \begin{minipage}{\dimexpr\linewidth-2.7cm-3.3cm}
    \includegraphics[width=\linewidth]{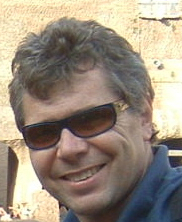}%
  \end{minipage}} 
{\bf Darwin G Caldwell} received the B.Sc. and Ph.D. degrees in robotics from University of Hull, Hull, U.K., in 1986 and 1990, respectively, and the M.Sc. degree 
in management from University of Salford, Salford, U.K., in 1996. He is the Director of Robotics with Istituto Italiano di Tecnologia, Genova, Italy. He is a Visiting/Honorary/Emeritus Professor with University of
Sheffield, the University of Manchester, and University of Wales, Bangor. His research interests include innovative actuators and sensors, haptic feedback, force augmentation exoskeletons, dexterous manipulators, humanoid robotics, 
biomimetic systems, rehabilitation robotics, and telepresence and teleoperation procedures. 
\par}

\end{multicols}
\end{document}